\documentclass[12pt]{article}

\ifx\pdfoutput\undefined
\usepackage[dvips,bookmarks]{hyperref}
\else
\usepackage{hyperref}
\fi
\hypersetup{colorlinks=false,bookmarksopen,bookmarksnumbered,citecolor=blue,
   pdfstartview=FitH}

\usepackage[dvips]{graphicx}
\usepackage{latexsym}
\usepackage{amssymb,amsfonts,amsmath}
\usepackage{graphicx} 
\usepackage{indentfirst}

 \usepackage{bbm}

\parskip=\medskipamount

\newcommand {\cD}{{\cal D}}
\newcommand {\cE}{{\cal E}}
\newcommand {\cF}{{\cal F}}
\newcommand {\cG}{{\cal G}}

\newcommand {\cL}{{\cal L}}
\newcommand {\cM}{{\cal M}}
\newcommand {\cN}{{\cal N}}

\newcommand {\cQ}{{\cal Q}}
\newcommand {\cR}{{\cal R}}
\newcommand {\cS}{{\cal S}}
\newcommand {\cT}{{\cal T}}

\newcommand {\cV}{{\cal V}}
\newcommand {\cW}{{\cal W}}

\newcommand {\cY}{{\cal Y}}

\newcommand{\bms}{{{\bm s}}}
\newcommand{\bmS}{{{\bm S}}}

\newcommand{\mub}{{{\bar{\mu}}}}
\def\a{\alpha}
\def \bi{\bibitem}

\def\b{\beta}

\def\d{\delta}
\def\e{\epsilon}
\def\f{\phi}
\def\g{\gamma}
\def\G{\Gamma}

\def\l{\lambda}
\def\m{\mu}

\def\o{\omega}

\def\q{\theta}

\newcommand{\qb}{{\bar{\theta}}}

\def\r{\rho}
\def\s{\sigma}

\def\x{\xi}
\def\z{\zeta}
\def\D{\Delta}
\def\F{\Phi}
\def\J{\Psi}
\def\L{\Lambda}
\def\O{\Omega}

\def\S{\Sigma}
\def\U{\Upsilon}
\def\X{\Xi}

\def\rd{{\rm d}}
\def\ri{{\rm i}}
\def\re{{\rm e}}

\newcommand{\ad}{{\dot{\alpha}}}                           
\newcommand{\bd}{{\dot{\beta}}}                            
\newcommand{\ve}{\varepsilon}                            
\newcommand{\cDB}{{\bar\cD}}                            
\newcommand{\DB}{\bar{D}}

\newcommand{\pa}{\partial}                           
\newcommand{\hf}{\frac12}

\newcommand{\vf}{\varphi}

\newcommand{\be}{\begin{equation}}
\newcommand{\ee}{\end{equation}}
\newcommand{\bea}{\begin{eqnarray}}
\newcommand{\eea}{\end{eqnarray}}
\newcommand{\non}{\nonumber}
\newcommand{\1}{{\underline{1}}}
\newcommand{\2}{{\underline{2}}}

\newcommand{\bm}[1]{\mbox{\boldmath$#1$}}

\def\double #1{#1{\hbox{\kern-2pt $#1$}}}

\newcommand{\gd}{{\dot\g}}
\newcommand{\dd}{{\dot\d}}

\newcommand{\ts}{{\tilde{\s}}}

\newcommand{\CD}{{\nabla}}
\newcommand{\CDB}{{\bar{{\nabla}}}}

\newcommand{\sba}{{\bar{\s}}}

\newcommand{\teb}{{\bar{\theta}}}

\renewcommand{\(}{\left(}
\renewcommand{\)}{\right)}

\newcommand{\vfb}{{\bar{\varphi}}}

\newcommand{\xb}{{\bar{\x}}}
\newcommand{\lb}{{\bar{\l}}}

\begin{document}

\begin{titlepage}

\begin{flushright}
July, 2008\\
\end{flushright}
\vspace{5mm}

\begin{center}
{\Large \bf  Field theory in 4D $\bm{\cN=2}$ 
conformally flat superspace}
\end{center}

\begin{center}

{\large  
Sergei M. Kuzenko\footnote{{kuzenko@cyllene.uwa.edu.au}}
and 
Gabriele Tartaglino-Mazzucchelli\footnote{gtm@cyllene.uwa.edu.au}
} \\
\vspace{5mm}

\footnotesize{
{\it School of Physics M013, The University of Western Australia\\
35 Stirling Highway, Crawley W.A. 6009, Australia}}  
~\\

\vspace{2mm}

\end{center}
\vspace{5mm}

\begin{abstract}
\baselineskip=14pt
Building on the superspace  formulation for 
four-dimensional $\cN=2$ matter-coupled supergravity 
developed in \cite{KLRT-M},  we elaborate upon a general setting for field theory 
in  $\cN=2$ conformally flat superspaces, 
and concentrate specifically on the case of
anti-de Sitter (AdS) superspace.
We demonstrate, in particular,  that associated with the $\cN=2$ AdS supergeometry 
is a unique vector multiplet
such that the corresponding  covariantly chiral field strength $\cW_0$
is constant, $\cW_0=1$.
This multiplet proves to be intrinsic in the sense that it encodes all the information 
about the $\cN=2$ AdS supergeometry in a conformally flat frame.
Moreover, it
emerges as a building block in the  construction of various supersymmetric actions.
Such a vector multiplet, which can be identified with one of the two compensators of 
$\cN=2$ supergravity,   also naturally occurs
for arbitrary conformally flat superspaces. 
An explicit superspace reduction $\cN=2 \to \cN=1$ 
is performed for the action principle in  general conformally flat $\cN=2$ backgrounds,
and examples of such reduction are given.
\end{abstract}
\vspace{1cm}

\vfill
\end{titlepage}

\newpage
\renewcommand{\thefootnote}{\arabic{footnote}}
\setcounter{footnote}{0}

\tableofcontents{}
\vspace{1cm}
\bigskip\hrule


\section{Introduction}
\setcounter{equation}{0}

Recently, we have developed the superspace formulation 
for four-dimensional $\cN=2$ matter-coupled supergravity \cite{KLRT-M}, 
extending the earlier construction for 5D $\cN=1$  
supergravity \cite{KT-Msugra1,KT-Msugra3}.
The locally supersymmetric action proposed in \cite{KLRT-M}
has a striking  similarity with  the chiral action 
in 4D $\cN=1$ supergravity \cite{Zumino78,SG}
(see also \cite{GGRS,BK} for reviews).
The  action functional proposed in \cite{KLRT-M}
can be written in the form:
\bea
S&=&
\frac{1}{2\pi} \oint (u^+ \rd u^{+})
\int \rd^4 x \,{\rm d}^4\q{\rm d}^4{\bar \q}\,
\cE\, \frac{\cL^{++}}{\cS^{++} \widetilde{\cS}^{++}}~, 
\qquad 
u^+_i \cD^{i}_\a \cL^{++} = u^+_i {\bar \cD}^{i}_\ad \cL^{++} =0~,~~~~
\label{InvarAc}
\eea
with $\cS^{++}(u^+):=\cS^{ij}u^+_i u^+_j$
and $\widetilde{\cS}^{++}(u^+):={\bar \cS}^{ij}u^+_i u^+_j$.
Here  $\cE^{-1}= {\rm Ber}(\cE_{\underline{A}}{}^{\underline{M}})$, 
where $\cE_{\underline{A}}{}^{\underline{M}}$ is
the (inverse) vielbein  appearing in the superspace covariant derivatives,
$\cD_{\underline{A}} =(\cD_{{a}}, \cD_{{\a}}^i,\cDB^\ad_i)$,
and $\cS^{ij}$ and ${\bar \cS}^{ij}$
are special irreducible  components of the torsion (see  Appendix A for more detail).
The Lagrangian $\cL^{++}(u^+)$ is a holomorphic homogeneous 
function of second degree with respect to auxiliary  
isotwistor variables $u^+_i \in {\mathbb C}^2 \setminus \{0\}$, which 
are introduced in addition to the superspace coordinates.
The total  measure in (\ref{InvarAc}) includes a contour integral 
 in the auxiliary isotwistor space.

Let us now recall the well-known chiral action \cite{Zumino78,SG}
in 4D $\cN=1$ old minimal ($n=-1/3$) supergravity
\cite{WZ-s,old}:
\bea
S_{\rm chiral}&=&
\int \rd^4 x \,{\rm d}^2\q{\rm d}^2{\bar \q}\,
E\, \frac{ L_{\rm c}}{R}~, 
\qquad 
{\bar \nabla}_\ad L_{\rm c} =0~.
\label{InvarAcN=1}
\eea
Here $E^{-1}$  is the superdeterminant of the (inverse) vielbein  
$E_A{}^M$ that enters the corresponding superspace covariant derivatives
$\nabla_{{A}} =(\nabla_{{a}}, \nabla_{{\a}},{\bar \nabla}^\ad)$,  
and $R$ is the chiral scalar component of the torsion
(following the notation of \cite{BK}).   
The action is generated by a covariantly chiral scalar Lagrangian $L_{\rm c}$.

The similarity between (\ref{InvarAc}) and (\ref{InvarAcN=1}) is at least twofold. First of all, 
each  action 
involves  integration over the corresponding {\it full} superspace. 
Secondly, the Lagrangians in (\ref{InvarAc}) and (\ref{InvarAcN=1}) 
obey covariant constraints which enforce 
$\cL^{++}$ and $L_{\rm c}$
to depend on {\it half} of the corresponding superspace Grassmann variables.
The latter property is of crucial importance. 
It indicates that there should exist a covariant way 
to rewrite each action as an integral over a submanifold 
of the full superspace such that the number of its fermionic directions is
half of the number of such variables 
in the full superspace (i.e. two in the $\cN=1$ case and four if $\cN=2$).
In the $\cN=1$ case, such a reformulation is well-known.
Using the chiral supergravity prepotential \cite{SG}, the action (\ref{InvarAcN=1}) 
can be rewritten as an integral over the chiral subspace 
of the curved superspace, see also \cite{GGRS,BK} for reviews
(a somewhat more exotic scheme
is presented in \cite{WB}). What about the $\cN=2$ case?
There are numerous reasons to expect
that the action (\ref{InvarAc}) can be reformulated 
as an integral over an $\cN=1$ subspace of the curved $\cN=2$ superspace.
In particular, this idea is natural from the point of view
of the projective superspace approach \cite{KLR,LR} to rigid 
$\cN=2$ superymmetric theories (the supergravity formulation
given in \cite{KLRT-M} can be viewed to be a curved  
projective superspace).
We hope to give a detailed elaboration of this proposal elsewhere.\footnote{In the 
1980s, there appeared a series of papers \cite{GKLR,Labastida} devoted to 
projecting special off-shell $\cN=2$ supergravity theories into $\cN=1$
superspace. Specifically: (i) Refs. \cite{GKLR} dealt with 
the standard $40+40$ formulation \cite{FV} for $\cN=2$
Poincar\'e supergravity realized in $\cN=2$ superspace \cite{BS,Howe}; 
and (ii) Ref.  \cite{Labastida} was concerned with  $\cN=2$ conformal supergravity 
realized in $\cN=2$ superspace in \cite{Howe}. Since off-shell formulations
for general matter couplings in $\cN=2$ supergravity 
were not available at that time, applications of  \cite{GKLR,Labastida} 
were rather limited. We hope that the progress achieved in \cite{KLRT-M}
should revitalize  the approaches pursued in  
\cite{GKLR,Labastida}.} 
Here we only provide partial supportive evidence  
by considering arbitrary conformally flat $\cN=2$ superspaces, 
including a maximally symmetric supergravity 
background -- 4D $\cN=2$ anti-de Sitter superspace. 

Unlike the case of simple anti-de Sitter supersymmetry (AdS)
in four dimensions,\footnote{The structural aspects  of 4D $\cN=1$ AdS superspace
and corresponding field representations were thoroughly studied
in \cite{IS} (see also \cite{Keck,Zumino} for earlier work).} 
field theory in  the $\cN=2$ AdS superspace is practically 
terra incognita.\footnote{The necessity of having an adequate superspace setting 
for $\cN=2$ AdS supersymmetry became apparent in \cite{GKS}
where {\it off-shell} higher spin supermultiplets  with $\cN=2$ 
AdS supersymmetry were constructed. These $\cN=2$ supermultiplets 
were realized in \cite{GKS} as field theories in the $\cN=1$ AdS superspace, 
by making use of the dually equivalent formulations for $\cN=1$ supersymmetric 
higher spin theories previously developed in  \cite{KS}. However,  their off-shell $\cN=2$ 
structure clearly hinted at the existence of a manifestly supersymmetric formulation in 
the  $\cN=2$ AdS superspace.  Some progress toward constructing 
such a formulation has been made in \cite{SS}.} 
In the case of the $\cN=2$ Poincar\'e supersymmetry in four dimensions, 
there exist two universal schemes to formulate general off-shell supersymmetric 
theories: the harmonic superspace \cite{GIKOS,GIOS} and the projective superspace 
\cite{KLR,LR}. To the best of our knowledge, no thorough analysis has been given in the literature 
regarding an extension of these approaches to the anti-de Sitter supersymmetry. 
One of the goals of the present paper is to fill this gap.

Before turning to the technical part of this paper, 
a  comment is in order. 
The action (\ref{InvarAc}) is equivalent to that originally 
given in \cite{KLRT-M}. The latter 
looks like 
\bea
S&=&
\frac{1}{2\pi} \oint (u^+ \rd u^{+})
\int \rd^4 x \,{\rm d}^4\q{\rm d}^4{\bar \q}\,
\cE\, \frac{\cW{\bar \cW}\,\cL^{++}}{(\S^{++})^2}~, 
\label{InvarAc-sWeyl-cov}
\eea
where $\cW$ is the covariantly chiral field strength, ${\bar \cD}^\ad_i \cW=0$, 
of an Abelian vector multiplet such that $\cW$ is everywhere non-vanishing,  and  
\be
\S^{++}(u^+):=\S^{ij}u^+_i u^+_j~, \quad
\S^{ij} =\frac{1}{4}\big(\cD^{\g(i}\cD_\g^{j)}+4\cS^{ij}\big)\cW
=\frac{1}{4}\big(\cDB_\gd^{(i}\cDB^{ j) \gd}+ 4\bar{\cS}^{ij}\big)\bar{\cW}~.~~
\ee
Unlike (\ref{InvarAc}), a notable feature of (\ref{InvarAc-sWeyl-cov})
is that it is manifestly super-Weyl invariant \cite{KLRT-M}.
The $\cN=1$ action  (\ref{InvarAcN=1}) can also be rewritten in a 
manifestly super-Weyl invariant form:
\bea
S&=&
\int \rd^4 x \,{\rm d}^2\q{\rm d}^2{\bar \q}\,
E\, \frac{ {\bar \J} \,\cL_{\rm c}}{\S}~, 
\qquad 
\S= -\frac{1}{4} \big({\bar \nabla}^2 -4R \big) 
{\bar \J} ~, \qquad {\bar \nabla}_\ad \J =0~.
\eea
Here $\J$ is a covariantly chiral scalar superfield
required  to be everywhere non-vanishing but otherwise arbitrary.

This paper is organized as follows. In section 2, after a brief review 
of the differential geometry of the 4D $\cN=2$ AdS superspace, AdS${}^{4|8}$, 
we elucidate  the structure of $\cN=2$ AdS Killing supervectors, and then 
introduce projective supermultiplets living in AdS${}^{4|8}$. 
In section 3, the manifestly supersymmetric action in AdS${}^{4|8}$
is reduced to $\cN=1$ superspace, and then several models 
for hypermultiplets, tensor and vector multiplets are  considered.
Section 4 begins with a general discussion of $\cN=2$ conformally flat superspaces. 
We then realize the  $\cN=2$ AdS superspace as locally conformal flat, 
work out the tropical prepotential for the intrinsic vector multiplet, 
and explicitly compute the  $\cN=2$ AdS Killing supervectors.
In section 5, the action  (\ref{InvarAc}) in an arbitrary
 conformally flat $\cN=2$ superspace is reduced to $\cN=1$ superspace.
 As applications of this reduction, we consider several models for massive hypermultiplets 
 in AdS${}^{4|8}$
 and vector multiplets in the conformally flat superspace. Final comments and conclusions 
 are given in section 6. The paper also contains four technical appendices. 
 Appendix A is devoted to a short review of the superspace geometry 
 of $\cN=2$ conformal supergravity following \cite{KLRT-M}. 
 In Appendix B, we elaborate upon the projective-superspace description  
 of Abelian vector multiplets 
 in conformal supergravity  (along with some properties previously presented 
 in \cite{KLRT-M},  new results  are included in this appendix).
Appendix C is devoted to a mini-review of the geometry of $\cN=1$ AdS superspace
and the corresponding Killing supervectors, following \cite{BK}. Finally, 
Appendix D presents a summary of the stereographic projection for 
$d$-dimensional AdS spaces.

\section{$\cN=2$ anti-de Sitter supergeometry}
\setcounter{equation}{0}

The superspace geometry, which is quite compact to use and, at the same time,
perfectly suitable to describe 
4D $\cN=2$ conformal supergravity and covariant 
projective matter supermultiplets, was presented in 
\cite{KLRT-M} (see  Appendix A for a concise  review); 
its connection to Howe's formulation for conformal supergravity \cite{Howe} 
is discussed in \cite{KLRT-M}.
In such a  setting,  the 4D $\cN=2$ AdS superspace 
$$
{\rm AdS}^{4|8} = \frac{{\rm OSp}(2|4)}{{\rm SO}(3,1) \times {\rm SO} (2)}
$$
corresponds to a geometry with covariantly constant torsion:\footnote{Compare with the case 
of 5D $\cN=1$ anti-de Sitter superspace \cite{KT-M}.}
\be
\cW_{\a\b}=\cY_{\a\b}=0~, 
\qquad \cG_{\a \bd}=0~, \qquad 
\cD^i_\a \cS^{kl} = {\bar \cD}^i_\ad \cS^{kl}=0~.
\label{AdS-geometry1}
\ee
The integrability condition for these constraints 
is $[\cS, \cS^\dagger ]=0$, with $\cS= (\cS^i{}_j)$,
and hence 
\be 
{\cS}^{ij} = q \, {\bm S}^{ij}~, \qquad  
 \overline{{\bm S}^{ij}}= {\bm S}_{ij}~,\qquad \quad |q|=1~,
\label{AdS-geometry2}
\ee
where   $q $ is a constant parameter.
By applying a rigid U(1) phase transformation to the covariant derivatives, 
$\cD_\a^i \to q^{-1/2} \cD_\a^i$,  
one can set $q=1$.
This choice will be assumed in what follows.

The covariant derivatives of the 4D $\cN=2$ AdS superspace  
form the following algebra:\begin{subequations}\bea\{\cD_\a^i,\cD_\b^j\}&=&
4{\bmS}^{ij}M_{\a\b}
+2 \ve_{\a\b}\ve^{ij}\bmS^{kl}J_{kl}~,
\qquad
\{\cD_\a^i,\cDB^\bd_j\}=
-2\ri\d^i_j(\s^c)_\a{}^\bd\cD_c
~,~~~
\label{AdS-N2-1}
\\
{[}\cD_a,\cD_\b^j{]}&=&
{\ri\over 2} ({\s}_a)_{\b\gd}\bmS^{jk}\cDB^\gd_k~,
\qquad \qquad \qquad \quad ~~
[\cD_a,\cD_b]= - \bmS^2
M_{ab}~,
\label{AdS-N2-2}
\eea
\end{subequations} 
with $\bmS^2 := \hf\bmS^{kl}\bmS_{kl}$.
These anti-commutation relations follow from 
(\ref{acr1}--\ref{acr3}) by choosing the torsion 
to be covariantly constant. 

In accordance with the general supergravity definitions given in Appendix A, 
the covariant derivatives include an appropriate  SU(2) connection, 
see eq. (\ref{SU(2)connection}).  It follows from (\ref{AdS-N2-1}), however, 
that the corresponding curvature is generated by a U(1) subgroup of SU(2). 
Therefore, one can gauge away most of the SU(2) connection except its U(1)
part corresponding to the generator  $  {\bm S}^{kl}J_{kl}$ 
\be
\F_A{}^{kl}J_{kl} \quad\longrightarrow \quad \F_A\, {\bm S}^{kl}J_{kl}~.
\label{SU(2)conn}
\ee
In such a gauge, the torsion ${\bm S}^{ij}$ becomes constant, 
\be
{\bm S}^{ij} ={\rm const}~.
\ee
By applying a rigid SU(2) rotation to the covariant derivatives, 
we can always choose 
\be
{\bm S}^{\1 \2} =0~.
\label{S12}
\ee
This choice will be 
often used in what follows.

\subsection{$\cN=2$ AdS Killing supervectors: I}

In this subsection, we do not assume any particular coordinate frame 
for the AdS covariant derivatives $\cD_{\underline{A}}$.
In particular, we do not impose the gauge fixing (\ref{SU(2)conn}).

The isometry transformations of AdS$^{4|8}$ form the group  OSp(2$|$4). 
Their  explicit structure 
can be determined in a manner similar 
to the cases of 4D $\cN=1$ AdS superspace \cite{BK} and 5D $\cN=1 $
superspace \cite{KT-M}. In the infinitesimal case,  an isometry transformation 
is generated by a real supervector field 
$\x^{\underline{A}} \,\cE_{\underline{A}}$ 
such that the operator 
\bea
&\x:=\x^{\underline{A}} (z) \cD_{\underline{A}}
= \x^a\cD_a+\x^\a_i\cD_\a^i+\xb_\ad^i\cDB^\ad_i~
\eea
 enjoys the property
\bea
\big[\x+\hf \l^{cd}M_{cd}
+\l^{kl}J_{kl}
,\cD_{\underline{A}}\big] =0~,
\label{Super-K-eq-00}
\eea
for some real antisymmetric tensor $\l^{cd}(z)$ and 
real symmetric tensor $\l^{kl}(z)$, $\overline{ \l^{kl}}=\l_{kl}$.
The latter equation implies 
\be
\big[\x+\l^{kl}J_{kl},  \bmS^{ij }\big] 
= \big[\l^{kl}J_{kl},  \bmS^{ij }\big] =0~,
\ee
and hence $\l^{kl} \propto  \bmS^{kl}$. 
We therefore can replace (\ref{Super-K-eq-00}) with 
\bea
\big[\x+\hf \l^{cd}M_{cd}
+\r \bmS^{kl}J_{kl}
,\cD_{\underline{A}}\big] =0~,
\label{Super-K-eq}
\eea
for some real scalar $\r(z)$. 
The meaning of  eq. (\ref{Super-K-eq}) is that 
the covariant derivatives do not change under 
the combined infinitesimal transformation consisting of 
coordinate ($\x$), 
local Lorentz  ($\l^{cd}$) and 
local U(1) ($\r$) transformations. 
It turns out that eq. (\ref{Super-K-eq}) uniquely determines 
the parameters $\l^{cd}$ and $\r$ in terms of $\x$. 
The $\x^{\underline{A}} \,\cE_{\underline{A}}$  is called a Killing supervector field.
The set of all Killing supervector fields forms a Lie algebra, 
with respect to the standard Lie bracket,
isomorphic to that of the group  OSp(2$|$4). 

Eq. (\ref{Super-K-eq}) implies that 
the parameters $\x^{\underline{A}}$,  $\l^{cd}$ and $\r$
are constrained as follows:
\begin{subequations} 
\bea
\cD_\a^i\x^\b_j
-\r \bmS^{i}{}_j\d_\a^\b
-\hf\l_\a{}^\b\d^i_j &=&0~,
\label{SK-1-1}
\\
\cDB^\ad_i\x^\b_j -{\ri\over 2}\bmS_{ij}\x^{\ad\b}
&=&0~,
\label{SK-1-2}
\\
\cDB^\ad_i\x^b
+2\ri\x^\b_i(\s^b)_\b{}^\ad&=&0~,
\label{SK-1-3}
\\
\cD_\a^i\l^{cd} - 4\bmS^{ij}\x^\b_j(\s^{cd})_{\a\b} &=&0~,
\label{SK-1-4}
\\
\cD_\a^i\r - 2\x_\a^i
&=&0 ~.
\label{SK-1-5}
\eea
\end{subequations} 
Note that eq. (\ref{SK-1-1}) is equivalent to
\bea
\cD_\g^k\x^\g_{k}
=\cD_{(\a}^{(i}\x_{\b)}^{j)}=0
~,
\qquad
2\r \bmS^{ij}
+\cD^{\g(i}\x_\g^{j)} =0
~,
\qquad
\l_{\a\b}=
\hf\cD_{(\a}^k\x_{\b) k}
~.
\label{SK-1-1-4}
\eea
Equation (\ref{SK-1-2}) is equivalent to
\bea
\cDB^\ad_k\x^{\b k}=0~,
\qquad
\cDB^\ad_{(i}\x^\b_{j)}
-{\ri\over 2} \bmS_{ij}\x^{\ad\b}
=0~.
\label{SK-2-2-2}
\eea
Equation (\ref{SK-1-3}) is equivalent to
\bea
\cDB^{(\ad}_i\x^{\gd)\g} =0
~,
\qquad
\cDB_{\gd i}\x^{\gd\g}
-8\ri\x^\g_{i} =0~.
\label{SK-2-3-2}
\eea
Equation (\ref{SK-1-4}) is equivalent to
\bea
\cDB^\ad_i\l^{\g\d} =0~, \qquad
\cD_{(\a}^i\l_{\g\d)}=0~,
\qquad
\cD^{\g i}\l_{\g\d}
+6\bmS^{ij}\x_{\d j}
=0~.
\label{SK-1-4-3}
\eea
It is also worth  noting that the above equations imply
\bea
\cD_{(a}\x_{b)}=0~
\eea
which is a natural generalization of the standard equation for Killing vectors.

Similar to the case of 5D $\cN=1$ AdS superspace \cite{KT-M}, 
all the components $\x^{\underline{A}}$ can be expressed 
in terms of the scalar parameter $\r$ as follows:
\bea
&&\x^\a_i=\hf
\cD^\a_i\r~,\qquad
\x_{\a\bd}={\ri\over 2\bmS^2}\bmS_{ij}\cD_\a^{i}\cDB_\bd^{j}\r~,
\qquad
\l_{\a\b}={1 \over 4}\cD_{\a}^k\cD_{\b k}\r~.
\eea
The latter obeys a number of constraints including
\bea
\Big(\cD^{\g i}\cD_\g^j+4 \bmS^{ij}\Big)\r =0~,\qquad
\Big(\cD_\a^{i}\cDB_\bd^{j}-{1\over 2\bmS^2}\bmS^{ij}\bmS_{kl}\cD_\a^{k}\cDB_\bd^{l}\Big)\r =0~,
\eea
and hence
\bea
\cD_a\r=0~.
\eea

\subsection{$\cN=1$ reduction} 

It is of interest to work out  $\cN=1$ components of the $\cN=2$ Killing
supervectors,  as well as of covariant $\cN=2$ supermultiplets. 
Given a tensor superfied  $U(x,\q_i,\qb^i)$ in $\cN=2$ 
AdS superspace, we introduce
its $\cN=1$ projection
\bea
U|:=U(x,\q_i,\qb^i)|_{\q_\2={\bar \q}^\2=0}~
\eea
in a {\it special coordinate system} to be specified below.
For the covariant derivatives 
\be
\cD_{\underline{A}}=\cE_{\underline{A}}{}^{\underline{M}}\pa_{\underline{M}}
+\hf\O_{\underline{A}}{}^{bc}M_{bc}+\F_{\underline{A}} \bmS^{kl}J_{kl}~,
\label{cov-der-U(1)-rep}
\ee
the projection is defined according to
\bea
\cD_{\underline{A}}|:=\cE_{\underline{A}}{}^{\underline{M}}|\pa_{\underline{M}} |
+\hf\O_{\underline{A}}{}^{bc}|M_{bc}+\F_{\underline{A}} | \bmS^{kl}J_{kl}~.
\eea
Here the first term on the right, 
$\cE_{\underline{A}}{}^{\underline{M}}|\pa_{\underline{M}}|$, 
includes the partial derivatives with respect to  the local coordinates of 
$\cN=2$ AdS superspace.

With the choice
$\bmS^{\1\2}=0$, as in eq. (\ref{S12}),  
it follows from  (\ref{AdS-N2-1}) and (\ref{AdS-N2-2}) that
\bea
\{\cD_\a^\1,\cD_\b^\1\}=
4{\bmS}^{\1\1}M_{\a\b}
~,~~
\{\cD_\a^\1,\cDB^\bd_\1\}=
-2\ri(\s^c)_\a{}^\bd\cD_c
~,
~~
{[}\cD_a,\cD_\b^\1{]}=
{\ri\over 2}({\s}_a)_{\b\gd}\bmS^{\1\1}\cDB^\gd_\1
~.~~~~~~
\label{D^1-alg}
\eea
Therefore,  
the operators $(\cD_a,\,\cD_\a^\1,\,\cDB^\ad_\1)$ form a closed  algebra
which is in fact isomorphic to that of the covariant derivatives   for 
$\cN=1$ AdS superspace with
\bea
\mub=-\bmS^{\1\1}~,
\label{S11-mu}
\eea
see Appendix C.
Note also that no U(1) curvature is  present in (\ref{D^1-alg}).

We use the freedom to perform general coordinate, local Lorentz and  U(1) 
transformations  to choose the gauge
\bea
\cD^\1_\a|=\CD_\a~, \qquad \cDB^\ad_\1|=\CDB^\ad~,
\eea
with $\nabla_A = (\nabla_a , \CD_\a ,\CDB^\ad)$ the covariant derivatives
for $\cN=1$ anti-de Sitter superspace (see Appendix C).
In such a coordinate system,
the operators $\cD_\a^\1|$ and $\cDB_{\ad \1}|$ do not involve any 
partial derivatives with respect to $\q_\2$ and ${\bar \q}^\2$, 
and therefore, for any positive integer $k$,  
it holds that $\big( \cD_{\hat{\a}_1} \cdots  \cD_{\hat{\a}_k} U \big)\big|
= \cD_{\hat{\a}_1}| \cdots  \cD_{\hat{\a}_k}| U|$, 
where $ \cD_{\hat{\a}} :=( \cD_\a^\1, {\bar \cD}^\ad_\1)$ and $U$ is a tensor superfield.

Given an arbitrary $\cN=2$ AdS Killing supervector $\x$, 
we consider its $\cN=1$ projection
\bea
\x|=\l^a\CD_a+\l^\a\CD_\a+\lb_\ad\CDB^\ad+\ve^\a\cD_\a^\2|+\bar{\ve}_\ad\cDB^\ad_\2|~,
\eea
where we have defined 
\bea
&&\l^{a}:=\x^{a}|~, \quad 
\l^\a:=\x^\a_\1|~,\quad 
\lb_\ad:=\xb_\ad^\1|~,\qquad 
\ve^\a:=\x^\a_\2|~, \quad \bar{\ve}_\ad:=\x_\ad^\2|
~.
\eea
We also introduce the projections of the parameters $\l_{ab} $ and $\r$:
\bea
&&\o_{ab}:=\l_{ab}|~,\qquad
\ve:=\r|~.
\eea
Now, the OSp$(2|4)$ transformation law of a tensor superfield $U$, 
\be
\d U =  \Big( \x+\hf \l^{cd}M_{cd}
+\r \bmS^{kl}J_{kl} \Big) U~, 
\ee
turns into 
\bea
\d U|&=&\Big(\l^a\CD_a+\l^\a\CD_\a+\lb_\ad\CDB^\ad+\hf\o^{ab}M_{ab} \Big) U| \non \\
&+& 
\Big( \ve^\a( \cD_\a^\2 U)| 
+\bar{\ve}_\ad (\cDB^\ad_\2 U)|\Big) 
~-~\ve(
{\bar \m}J_{\1\1}+
\m
J_{\2\2})U|~,
~~~~~~
\label{deltaU|}
\eea
where we have made use of (\ref{S11-mu}).
It can be shown that $\L =\l^a\CD_a+\l^\a\CD_\a+\lb_\ad\CDB^\ad$ 
is an $\cN=1$ AdS Killing supervector (see Appendix C),
and the variation in the first line of  (\ref{deltaU|}) is the  infinitesimal OSp$(1|4)$
transformation generated by $\L$.
The parameters $\ve^\a,\,\bar{\ve}_\ad$ and $\ve$
generate the second supersymmetry and U(1) transformations.
It can be shown, using eqs. (\ref{SK-1-1-4})--(\ref{SK-1-4-3}), 
that they obey the constraints \cite{GKS}
\bea
&&\ve^\a=\hf \CD^{\a}\ve~, \qquad
\CD_\a\CDB^\ad\ve=0~, \qquad 
\big(\CD^2-4\bar{\mu}\big)\ve=0~.
\label{epsilon}
\eea

\subsection{Projective supermultiplets in AdS$^{4|8}$}

General matter couplings in 4D $\cN=2$ supergravity can be described 
in terms of covariant projective supermultiplets \cite{KLRT-M}.
Here we briefly introduce such multiplets in the case of $\cN=2$ 
AdS superspace, and then work out their reduction to $\cN=1$ 
superfields.

In the superspace AdS$^{4|8}$, a  projective supermultiplet of weight $n$,
$\cQ^{(n)}(z,u^+)$, 
is defined to be a scalar superfield that
lives on  AdS$^{4|8}$,
is holomorphic with respect to
the isotwistor variables $u^{+}_i $ on an open domain of
${\mathbb C}^2 \setminus  \{0\}$,
and is characterized by the following conditions:\\
${}\quad$(1) it obeys the covariant analyticity constraints\footnote{In the rigid 
supersymmetric case, 
constraints of the form (\ref{ana}) in isotwistor superspace
${\mathbb R}^{4|8}\times {\mathbb C}P^1$  
were first introduced by Rosly 
\cite{Rosly},   and later by the harmonic \cite{GIKOS} 
and projective \cite{KLR,LR} superspace practitioners.}  
\be
\cD^+_{\a} \cQ^{(n)}  = {\bar \cD}^+_{\ad} \cQ^{(n)}  =0~, \qquad 
\cD^+_\a := u^+_i \cD^{i}_\a ~, \quad 
{\bar \cD}^+_\ad := u^+_i {\bar \cD}^{i}_\ad 
~;
\label{ana}
\ee
${}\quad$(2) it is  a homogeneous function of $u^+$
of degree $n$, that is,
\be
\cQ^{(n)}(z,c\,u^+)\,=\,c^n\,\cQ^{(n)}(z,u^+)~, \qquad c\in \mathbb{C}\setminus \{0\}~;
\label{weight}
\ee
${}\quad$(3)  the infinitesimal OSp$(2|4)$ transformations act on $\cQ^{(n)}$
as follows:
\bea
\d_\x \cQ^{(n)}
&=& \Big( \x^{{a}} \cD_{{a}}+\x^\a_i\cD_\a^i+\xb_\ad^i\cDB^\ad_i +\r \bmS^{ij} J_{ij} \Big)\cQ^{(n)} ~,
\non \\
\bmS^{ij} J_{ij}  \cQ^{(n)}&=& -\frac{1}{(u^+u^-)} \Big(\bmS^{++} D^{--}
-n \, \bmS^{+-}\Big) \cQ^{(n)} ~, \qquad
\bmS^{\pm \pm } =\bmS^{ij}\, u^{\pm}_i u^{\pm}_j ~,
\label{harmult1}
\eea
with $D^{--}=u^{-i}\frac{\partial}{\partial u^{+ i}}$.
The transformation law (\ref{harmult1}) involves an additional two-vector,  $u^-_i$, 
which is only subject 
to the condition $(u^+u^-) := u^{+i}u^-_i \neq 0$, and is otherwise completely arbitrary.
Both  $\cQ^{(n)}$ and $\bmS^{ij} J_{ij}  \cQ^{(n)}$ are independent of $u^-$.

In the family of projective multiplets, one can introduce a generalized conjugation, 
$\cQ^{(n)} \to \widetilde{\cQ}^{(n)}$, defined as
\be
\widetilde{\cQ}^{(n)} (u^+)\equiv \bar{\cQ}^{(n)}\big(
\overline{u^+} \to 
\widetilde{u}^+\big)~, 
\qquad \widetilde{u}^+ = {\rm i}\, \s_2\, u^+~, 
\ee
with $\bar{\cQ}^{(n)}(\overline{u^+}) $ the complex conjugate of $\cQ^{(n)}(u^+)$.
It is easy to check that $\widetilde{\cQ}^{ (n) } (z,u^+)$ is a projective multiplet of weight $n$.
One can also see that
$\widetilde{\widetilde{\cQ}}{}^{(n)}=(-1)^n \cQ^{(n)}$,
and therefore real supermultiplets can be consistently defined when 
$n$ is even.
The $\widetilde{\cQ}^{(n)}$ is called the smile-conjugate of 
${\cQ}^{(n)}$.

It is natural to interpret the variables $u^+_i$ as homogeneous coordinates
for ${\mathbb C}P^1$. 
Due to the homogeneity condition (\ref{weight}), 
the projective multiplets $\cQ^{ (n) } (z,u^+)$ actually depend on 
a single complex variable $\z$  which is an  inhomogeneous local complex coordinate
for ${\mathbb C}P^1$. 
To describe the projective multiplets in terms of $\z$, 
one should replace $\cQ^{(n)}(z,u^+)$ with a new superfield 
$\cQ^{[n]}(z,\z) \propto \cQ^{(n)}(z,u^+)$, where $\cQ^{[n]}(z,\z) $ is  holomorphic 
with respect to  $\z$.
The explicit definition of $\cQ^{[n]}(\z)$ depends on the supermultiplet under 
consideration.  
One can cover  ${\mathbb C}P^1$  
by two open charts in which $\z$ can be defined, 
and the simplest choice is:
(i) the north chart characterized by $u^{+\1}\neq 0$;
(ii) the south chart with  $u^{+\2}\neq 0$.
Our consideration will be restricted to the north chart 
in which the variable 
$\z \in \mathbb C$ is defined as
\bea
&&u^{+i} =u^{+\1}(1,\z) =u^{+\1}\z^i ~,\qquad
\z^i=(1,\z)~, \qquad \z_i= \ve_{ij} \,\z^j=(-\z,1)~.
\label{north}
\eea
In this chart, we can choose
\bea
&&u^-_i =(1,0) ~, \qquad   \quad ~u^{-i}=\ve^{ij }\,u^-_j=(0,-1)~.
\eea

Before  discussing the possible types of $\cQ^{[n]}(\z)$, 
let us first  turn to the U(1) part of the transformation law (\ref{harmult1}).
The parameters $\bmS^{++}$
and $\bmS^{+-}$ in (\ref{harmult1})
can be represented as $\bmS^{++} =\big(u^{+\1}\big)^2 \Xi (\z)$
and $\bmS^{+-}= u^{+\1}\D(\z) $, where
\bea
\Xi (\z)&=& {\bmS}^{ij} \,\z_i \z_j
=  {\bmS}^{\1 \1 }\, \z^2 -2  {\bmS}^{\1 \2}\, \z
+ {\bmS}^{\2 \2} ~,
\qquad
\D(\z)= {\bmS}^{\1 i} \,\z_i
=  - {\bmS}^{\1 \1} \,\z + {\bmS}^{\1 \2}  ~.~~~~~~~
\label{K++K}
\eea
Now, let us introduce the major projective supermultiplet  $\cQ^{(n)}(z,u^+)$
and the corresponding superfields $\cQ^{[n]}(z,\z)$. 
In the case of  covariant arctic weight-$n$ hypermultiplets $\U^{(n)}(z,u^+)$ \cite{KLRT-M}, 
it is natural to define
\bea
\U^{(n)}(z,u^+)=(u^{+\1})^n \U^{[n]}(z,\z)~, \qquad 
\U^{ [n] } (z, \z) = \sum_{k=0}^{\infty} \U_k (z) \z^k~.
\label{arctic-n}
\eea
The corresponding U(1) transformation law is:
\bea
\r \bmS^{ij}J_{ij} \U^{ [n] } (\z)
 &=& \r \Big( \Xi (\z)\,\pa_\z +  n\,\D(\z) \Big) \U^{[n]}(\z)~. 
 \eea
The smile-conjugate of $ \U^{(n)}$ is called 
a covariant  {\it antarctic} weight-$n$ multiplet. In this case
\bea
\widetilde{\U}^{(n)} (z, u) &=&  (u^{+\2})^n\, \widetilde{\U}^{[n]} (z, \z)~, \qquad
\widetilde{\U}^{[n]} (z, \z) = \sum_{k=0}^{\infty} (-1)^k {\bar \U}_k (z)
\frac{1}{\z^k}~,
\label{antarctic1}
\eea
with $ {\bar \U}_k$ the complex conjugate of $\U_k$.
The  U(1) transformation  of $\widetilde{\U}^{[n]} (z, \z)~$ is as follows:
\bea
 \r \bmS^{ij}J_{ij}  \widetilde{\U}^{[n]} (\z)=  
 \frac{\r}{\z^n}\Big(  \X(\z) \,\pa_\z 
 + n\,\D (\z) 
 \Big) \Big(\z^n\,\widetilde{\U}^{(n)} (\z)\Big)~.
\label{antarctic2}
\eea
In the case of a real weight-$2n$ projective superfield $R^{(2n)}(z,u^+)$, 
it is natural to define
\bea
R^{(2n)}(z,u^+)&=&
\big({\rm i}\, u^{+\1} u^{+\2}\big)^n R^{[2n]}(z,\z) ~.
 \label{real-2n}
\eea
The  U(1) transformation  of $R^{[2n]}(z,\z) $ is:
\bea
&&\r \bmS^{ij}J_{ij} R^{[2n]} = 
\frac{\r}{\z^n} \Big(    \Xi (\z) \,\pa_\z +2n\,\D(\z)\Big)
 \Big( \z^n R^{[2n]}\Big)~.
\eea
There are two major types of superfields $R^{[2n]}(z,\z)$:
a real $O(2n)$-multiplet ($n=1,2,\dots$) 
\bea
H^{[2n]}(z,\z) &=&
\sum_{k=-n}^{n} H_k (z) \z^k~,
\qquad  {\bar H}_k = (-1)^k H_{-k} ~,
\label{o2n1}
\eea
and a tropical weight-$2n$ multiplet
\bea
U^{[2n]}(z,\z) &=&
\sum_{k=-\infty}^{\infty} U_k (z) \z^k~,
\qquad  {\bar U}_k = (-1)^k U_{-k} ~. 
\label{trop-nj}
\eea

If the projective supermultiplet  $\cQ^{(n)}(z,u^+)$
is described by $\cQ^{[n]}(z,\z) \propto \cQ^{(n)}(z,u^+)$,
then the covariant analyticity conditions (\ref{ana}) become
\begin{subequations}
\bea
\cD^+_{ \a} (\z) \, \cQ^{[n]} (\z) &=&0~,
\qquad \cD^+_{ \a} (\z) = -\cD^i_{ \a} \z_i
=\z \,\cD^{\1}_{ \a}  - \cD^{\2}_{ \a} ~,
\label{ana1}  \\
{\bar \cD}^{+ \ad} (\z) \, \cQ^{[n]} (\z) &=&0~,
\qquad {\bar \cD}^{+ \ad} (\z)
= {\bar \cD}^\ad_i \z^i = {\bar \cD}^\ad_{\1}  +\z{\bar \cD}^\ad_{\2}~,
\label{ana2}
\eea
\end{subequations}
and therefore
\bea
&& \cD^{\2}_{ \a}\cQ^{[n]} (\z) =\z \,\cD^{\1}_{ \a}\cQ^{[n]} (\z)~, ~~~
{\bar \cD}^\ad_{\2}\cQ^{[n]} (\z)=-{1\over\z}{\bar \cD}^\ad_{\1}\cQ^{[n]} (\z)~.
\label{ana3}
\eea
The differential operator $\x^\a_i\cD_\a^i+\xb_\ad^i\cDB^\ad_i$,
which enters the transformation law
(\ref{harmult1}),
acts on $\cQ^{[n]}(\z)$ as
\bea
\big(\x^\a_i\cD_\a^i+\xb_\ad^i\cDB^\ad_i\big) \cQ^{[n]}(\z)
=\big((\x^\a_\1+\z\x^\a_\2)\cD^\1_\a+\big(\xb^\1_\ad-{1\over\z}\xb^\2_\ad\big)\cDB^\ad_\1
\big)\cQ^{[n]}(\z)~.
\label{anaconda}
\eea

Let us impose the SU(2) gauge (\ref{SU(2)conn})
and  choose  $\bmS^{\1\2}=0$, as in eq. (\ref{S12}).
Then, eq. (\ref{anaconda})
implies that the $\cN=1$ projection of $\x    \cQ^{[n]} (\z) $ is 
\bea
\Big( \x    \cQ^{[n]} (\z) \Big) \Big| = 
\L \cQ^{[n]}(\z)\big| + 
\Big( \z\ve^\a \CD_\a-{1\over\z}\bar{\ve}_\ad\CDB^\ad\Big)
\cQ^{[n]}(\z)\big|~,
\eea
with $\x$ an arbitrary  $\cN=2$ AdS Killing supervector, and $\L$ 
the induced $\cN=1$ AdS Killing supervector. 
As a result, the $\cN=1$ projection of the transformation $\d_\x \U^{[n]}(\z)$ becomes 
\bea
\d_\x \U^{[n]}(\z) \big|&=& 
\L \U^{[n]}(\z)\big| \non \\
&+& 
\big( \z\ve^\a \CD_\a-{1\over\z}\bar{\ve}_\ad\CDB^\ad\big)
\U^{[n]}(\z)\big|
+ \ve  \big( \Xi (\z)\,\pa_\z +  n\,\D(\z) \big) \U^{[n]}(\z)|~,
\label{arctic2}
\eea
and similarly for $\d_\x \widetilde{\U}^{[n]}(\z) \big|$.
The $\cN=1$ projection of the transformation $\d_\x R^{[2n]}(\z)$  becomes 
\bea
\d_\x R^{[2n]}(\z) \big|&=& 
\L R^{[2n]}(\z)\big| \non \\
&+& 
\big( \z\ve^\a \CD_\a-{1\over\z}\bar{\ve}_\ad\CDB^\ad\big)
R^{[2n]}(\z)\big|
+
\frac{\ve}{\z^n}\Big(    \Xi (\z) \,\pa_\z +2n\,\D(\z)
 \Big) \big( \z^n R^{[2n]}\big| \big)~.~~~~~~~~~
\label{transform-law-real}
\eea
In the gauge chosen, 
the parameters $\Xi(\z)$ and $\D(\z)$  are:
\bea
\Xi (\z)&=& -{\bar \m} \, \z^2 - \m
~,
\qquad
\D(\z)= {\bar \m}\,\z~.
\eea

\section{Dynamics in AdS$^{4|8}$}
\setcounter{equation}{0}

In the case of $\cN=2$ anti-de Sitter space, the action (\ref{InvarAc})
becomes
\bea
S&=&
\frac{1}{2\pi} \oint (u^+ \rd u^{+})
\int \rd^4 x \,{\rm d}^4\q{\rm d}^4{\bar \q}\,
\cE\, \frac{\cL^{++}}{(\bmS^{++})^2 }~, 
\label{InvarAc-AdS}
\eea
where the Lagrangian $\cL^{++}(z,u^+)$ is a real weight-two projective supermultiplet.

It is worth giving two non-trivial examples of supersymmetric theories  in AdS$^{4|8}$.
${}$First, we consider a superconformal model
of arctic weight-one hypermultiplets $\U^+$ 
and their smile-conjugates $\widetilde{\U}^+$
described by the Lagrangian \cite{K-hyper1,K-hyper2}
\be
\cL^{++}_{\rm conf}   =   {\rm i} \, K(\U^+, \widetilde{\U}^+)~,
\label{conf-matter}
\ee
where the real function $K(\F^I, {\bar \F}^{\bar J}) $
obeys the homogeneity condition 
\bea
\F^I \frac{\pa}{\pa \F^I} K(\F, \bar \F) =  K( \F,   \bar \F)~.
\label{Kkahler2}
\eea

Our second example is the
non-superconformal model of  arctic weight-zero multiplets 
${\bf \U}  $ and their smile-conjugates
$ \widetilde{ \bf{\U}}$  described by the Lagrangian \cite{KT-M}
\bea
\cL^{++}_{\rm non-conf} = {\bm S}^{++}\,
{\bf K}({\bf \U}, \widetilde{\bf \U})~,
\label{non-conf-sigma}
\eea
with ${\bf K}(\F^I, {\bar \F}^{\bar J}) $ a real function 
which is not required to obey any 
homogeneity condition. 
The action is invariant under K\"ahler transformations of the form
\be
{\bf K}({\bf \U}, \widetilde{\bf \U})~\to ~{\bf K}({\bf \U}, \widetilde{\bf \U})
+{\bf \L}({\bf \U}) +{\bar {\bf \L}} (\widetilde{\bf \U} )~,
\ee
with ${\bf \L}(\F^I)$ a holomorphic function.

Throughout this section, the torsion $\bmS^{ij}$
is chosen to obey eq.  (\ref{S12}).

\subsection{Projecting the $\cN=2$ action  into $\cN=1$  superspace: I}

In this subsection, we reduce the $\cN=2$ supersymmetric action (\ref{InvarAc-AdS})
to the $\cN=1$ AdS superspace.

Without loss of generality, the integration contour  in   (\ref{InvarAc-AdS})
can be assumed to lie outside the north pole $u^{+i} \propto (0,1)$, 
and then we can use the complex variable $\z$ defined in the north chart, 
eq. (\ref{north}), to parametrize the projective  supermultiplets. 
Associated with the Lagrangian  $\cL^{++}(u^+)$ is  
the superfield $\cL(\z)$ defined as
\bea
\cL^{++}(u^+):=\ri u^{+\1}u^{+\2}\cL(\z)=\ri (u^{+\1})^2\z\,\cL(\z)~.
\eea
Similarly, associated with $\bmS^{++}(u^+) $ is the superfield ${\bm S}(\z)$ defined as
\bea
\bmS^{++}(u^+):=\ri (u^{+\1})^2\z\,{\bm S}(\z)~,
\qquad {\bm S}(\z)=\ri\(\mub\,\z+\mu\,{1\over\z}\)~.
\label{S(zeta)}
\eea

Let ${\cL}(\z)|$ denote the $\cN=1$ projection of the Lagrangian  $\cL(\z)$.
Then, 
the manifestly $\cN=2$ supersymmetric functional  
(\ref{InvarAc-AdS}) 
can be shown to be equivalent to the following action in AdS$^{4|4}$:
\bea
S = \oint_C \frac{\rd \z}{2\pi \rm i\z}
\int \rd^4 x \, {\rm d}^2\q{\rm d}^2{\bar \q}\,{E}\,{\cL}(\z)|~,
\qquad {E}^{-1}:={\rm Ber}({E}_A{}^M)~.
\label{InvarAc5-N=2-N=1}
\eea
While this form of the action will be derived 
in section 5,
here we only demonstrate that (\ref{InvarAc5-N=2-N=1}) is invariant under the OSp$(2|4)$ 
transformations. We note that the transformation law of $\cL(\z)$ is given by eq. 
(\ref{transform-law-real}) with $n=1$.  
It is obvious that (\ref{InvarAc5-N=2-N=1}) 
is manifestly invariant under the $\cN=1$ AdS transformations 
\bea
\d_\L\cL(\z)|=\L\cL(\z)|=\(\l^a\CD_a+\l^\a\CD_\a+\lb_\ad\CDB^\ad\)\cL (\z)|~.
\eea
The other transformations, which 
are generated by the parameters $\ve,\,\ve^\a,\,\bar{\ve}_\ad$ 
in (\ref{transform-law-real}),  
act on $\cL(\z)|$ as follows: 
\bea
\d_\ve {\cL}(\z)| =
\Big(\z\ve^\a\CD_\a-{1\over\z}\bar{\ve}_\ad\CDB^\ad \Big) \cL (\z)|
-\frac{\ve}{\z}\Big(  \big(\z^2\mub+\mu\big)\pa_\z -2\z\mub\Big) \big(\z
{\cL}(\z)|\big)~.~~~~~~
\label{var-ve-cL}
\eea
The corresponding variation of the action,
\bea
\d_\ve S&=&\oint_C \frac{\rd \z}{2\pi \rm i\z}
\int \rd^4 x \, {\rm d}^2\q{\rm d}^2{\bar \q}\,{E}\,\d_\ve{\cL}(\z)|
~,
\eea
can be transformed by 
integrating by parts the derivatives $\CD_\a,\,\CDB^\ad$ and $\pa_\z$.
This leads to
\bea
\d_\ve S&=&\oint_C \frac{\rd \z}{2\pi \rm i\z}
\int \rd^4 x \, {\rm d}^2\q{\rm d}^2{\bar \q}\,{E}\Big(-\z(\CD^\a\ve_\a)+{1\over\z}(\CDB_\ad\bar{\ve}^\ad)
+2\ve\big(\mub\z- \mu\frac{1}{\z}\big)
\Big){\cL}(\z) |=0
~,~~~~~~~~~
\non
\eea
where we have made use of the relations
\bea
\ve^\a={1\over 2}\CD^\a\ve~, \qquad
\CD^\a\ve_\a=2\mub\ve~.
\label{Useful-Killing}
\eea

\subsection{Free hypermultiplets, dual tensor multiplets and some generalizations} 

To get a better feeling of the sigma-models
(\ref{Kkahler2}) and (\ref{non-conf-sigma}), 
it is instructive to examine their simplest versions 
corresponding to free hypermultiplets. 

Consider the Lagrangian
\bea
\cL_{\rm conf}^{++}&=&{\rm i} \,\widetilde{\U}^+\U^+
\label{confhyper}
\eea
which describes the dynamics of a weight-one 
arctic hypermultiplet $\U^+$ and its smile-conjugate $\widetilde{\U}^+$.

We represent $\U^+(u^+) =u^{+\1}\U (\z)$, where $\U(\z)$ is given  
by a convergent Taylor series centered  at $\z=0$.
Then, the analyticity conditions (\ref{ana3}) imply
\bea
\U(\z)|&=&\F+\z\G+\sum_{k=2}^{+\infty}\z^k \U_k|~,
\qquad \CDB^\ad\F=0~, \qquad \(\CDB^2-4\mu\)\G=0~.
\eea
Here $\F$ and $\G$ are covariantly chiral and complex linear superfields, 
respectively, while the higher-order components $\U_k|$, with $k=2,3,\dots$,
are complex unconstrained  superfields. 
It is useful to recall that, in the $\cN=1$ AdS superspace, 
the chirality constraint $\CDB^\ad\F=0$ is equivalent to 
$\CDB^2\F=0$ \cite{IS}. Moreover, any complex scalar superfield $U$ can be uniquely 
represented in the form $U=\F +\G$, for some chiral $\F$ and complex linear $\G$
scalars \cite{IS} (see \cite{West} for a nice review of the $\cN=1$ AdS 
supermultiplets classified in \cite{IS}).

Then, evaluating the action  (\ref{InvarAc5-N=2-N=1}) with $\cL(\z)$
corresponding to  (\ref{confhyper}) gives
\bea
S_{\rm conf}&=&
\oint_C \frac{\rd \z}{2\pi \rm i\z}
\int \rd^4 x \, {\rm d}^2\q{\rm d}^2{\bar \q}\, {E}\,\widetilde{\U}(\z)|\U(\z)|
\non\\
&=&
\int \rd^4 x \, {\rm d}^2\q{\rm d}^2{\bar \q} \,{E}\Big(\bar{\F}\F-\bar{\G}\G
+\sum_{k=2}^{+\infty}(-1)^k\bar{\U}_k|\U_k|\Big)~.~~~~~
\eea
Integrating out the auxiliary superfields $\U_k|$, 
in complete analogy with the flat case \cite{G-RLRvUW}, reduces the action to 
\bea
S_{\rm conf}=
\int \rd^4 x \, {\rm d}^2\q{\rm d}^2{\bar \q}\, {E}\(\bar{\F}\F-\bar{\G}\G\)
~.~~~~~
\label{conf-act-1}
\eea
The first term in the action provides the standard (or {\it minimal}) off-shell description 
of $\cN=1$ massless scalar multiplet.  The second term describes the same 
multiplet on the mass shell, although it is realized in terms of 
 a complex scalar and its conjugate. The latter description is known 
 as the {\it non-minimal} scalar multiplet \cite{GS}.
 
The action (\ref{conf-act-1}) is manifestly $\cN=1$ supersymmetric. 
It is also invariant under the second SUSY and U(1) transformations which 
are generated by a real parameter $\ve$ subject to the constraints (\ref{epsilon}),
and have the form:
\bea
\d_\ve \F=-\(\bar{\ve}_\ad\CDB^\ad+\ve\mu\){\G}=-{1\over 4}(\CDB^2-4\mu)(\ve\G)
~, \qquad
\d_\ve \G=\(\ve^\a\CD_\a+\ve\mub\)\F
~.
\eea
The complex linear superfield $\G$ can be dualized\footnote{The existence of a duality 
between the minimal $(\F, \bar \F )$ and the non-minimal $(\G, \bar \G )$ formulations
for scalar multiplet became apparent after the foundational work of \cite{SG}, where these 
realizations were shown to occur as the compensators corresponding to the 
old minimal and non-minimal formulations, respectively,
for $\cN=1$ supergravity.}    
into a covariantly chiral scalar superfield $\Psi$,  $\CDB^\ad \J=0$,
by applying a superfield Legendre transformation \cite{LR2} (see \cite{GGRS,BK} for reviews)
to end up with
\bea
S^{({\rm dual})}_{\rm conf}=
\int \rd^4 x \, {\rm d}^2\q{\rm d}^2{\bar \q}\, {E}\(\bar{\F}\F+\bar{\Psi}\Psi\)
~.~~~~~
\label{conf-act-1-dual}
\eea
The second SUSY and U(1) invariance of this 
model is as follows:
\bea
\d_\ve\F=-{1\over 4}(\CDB^2-4\mu)(\ve\bar{\Psi})~,\qquad
\d_\ve\Psi={1\over 4}(\CDB^2-4\mu)(\ve\bar{\F})~.
\label{SUSY-chiral-chiral}
\eea

Now consider the Lagrangian
\bea
\cL_{\rm non-conf}^{++}=\frac{\bmS^{++}}{|\bmS |}
\widetilde{\bm\U}{\bm\U}~
\label{L-0-massless}
\eea
describing the  dynamics of
a weight-zero arctic multiplet  ${\bm\U}$ and its conjugate $\widetilde{\bm\U}$.
Upon reduction to the $\cN=1$ AdS superspace,
this system is described by the action
\bea
S_{\rm non-conf}= \frac{1}{|\m|}
\oint_C \frac{\rd \z}{2\pi \rm i\z}
\int \rd^4 x \, {\rm d}^2\q{\rm d}^2{\bar \q}\,
 {E}\,
{\bm S}(\z)
 \widetilde{\bm\U}(\z)|{\bm\U}(\z)| ~,
 \label{non-con-ac}
\eea
where ${\bm S}(\z)$ is given in eq. (\ref{S(zeta)}). 
The $\cN=1$ projection of ${\bm\U}(\z)$ has the form:
\bea
{\bm\U}|(\z) &=&{\bm\F}+\z{\bm\G}+\sum_{k=2}^{+\infty}\z^k {\bm\U}_k|~,
\qquad \CDB^\ad{\bm\F}=0~, \qquad\(\CDB^2-4\mu\){\bm\G}=0~,
\label{weight-zero}
\eea
with the scalar superfields ${\bm\U}_k|$, $k\geq2$, being complex unconstrained.
To perform the contour integral in (\ref{non-con-ac}),  it is useful to note that
\bea
\frac{1}{ |\mu|}{\bm S}(\z)
=\Big(1-{1\over\l}{1\over\z}\Big)\Big(1+\l\z\Big)~,\qquad \l:=\ri\,{{\mub\over|\mu|}}~.~~
\eea
We then can redefine the components of the arctic multiplet as
\bea
&&{\bm{\U}}'|:=\Big(1+\l\z\Big){\bm\U}|
={\bm\F}'+\z{\bm\G}'+\sum_{k=2}^{\infty} {\bm\U}_k' \z^k~,~~~
\non\\
&&{\bm\F}'={\bm\F}~,\qquad
{\bm\G}'={\bm\G}+\l{\bm\F}~,\qquad
{\bm\U}_k'={\bm\U}_k|+\l{\bm\U}_{k-1}|
\quad,~
k>1~.~~~~~~
\label{U'}
\eea
Here ${\bm\G}'$ obeys  a modified
linear constraint of the form:
\bea
-{1\over 4}\(\CDB^2-4\mu\){\bm \G}'&=&{\rm i} |\mu|
{\bm\F}~.
\label{CNM}
\eea
Such a constraint is typical of chiral--non-minimal multiplets \cite{DG}. 
The complex superfields ${\bm\U}'_k$ with $k>1$ are obviously unconstrained.
Now, the contour integral in (\ref{non-con-ac}) can easily  be performed,
and the auxiliary fields  integrated out, whence 
the  action $S_{\rm non-conf}$ becomes
\bea
S_{\rm non-conf}
=\int \rd^4 x \, {\rm d}^2\q{\rm d}^2{\bar \q}\, E\Big(\bar{{\bm\F}}{\bm\F}-\bar{{\bm\G}}'{\bm\G}'\Big)
~.
\label{S-CNM}
\eea
The second SUSY and U(1) transformations of this action are:
\bea
\d_\ve {\bm\F}&=&\ri\ve|\mu|{\bm\F}-\(\bar{\ve}_\ad\CDB^\ad+\ve\mu\){\bm\G}'=
-{1\over 4}(\CDB^2-4\mu)(\ve{\bm\G}')
~, \non\\
\d_\ve {\bm\G}'&=&\ri\ve|\mu|{\bm\G}'+\(\ve^\a\CD_\a+\ve\mub\){\bm\F}
~.
\label{SUSY-chiral-linear}
\eea
The generalized complex linear superfield
${\bm\G}'$, which is  constrained by (\ref{CNM}), 
can be dualized into a covariantly chiral scalar
${\bm\Psi}$, ${\bar \nabla}^\ad  {\bm\Psi} =0$,
to result with the following purely chiral action: 
\bea
S_{\rm non-conf}^{({\rm dual})}&=&
\int \rd^4 x \, {\rm d}^2\q{\rm d}^2{\bar \q}\, E\Big(
\bar{{\bm\F}}{\bm\F}
+\bar{{\bm\Psi}}{\bm\Psi}
-\ri\frac{\bar \mu}{|\mu |}{\bm\Psi}{\bm\F} +\ri\frac{\mu}{| \mu |} {\bar {\bm\Psi}} {\bar{\bm\F} }
\Big)
~.
\label{S-CC}
\eea
In a flat superspace limit, $\m \to 0$, the last two terms in (\ref{S-CC}) 
will drop out.
The second SUSY and U(1) transformations of the model (\ref{S-CC}) 
coincide, modulo a simple re-labeling of the chiral variables, with (\ref{SUSY-chiral-chiral}).

The difference between the hypermultiplet models 
(\ref{confhyper}) and (\ref{L-0-massless})
can naturally be  understood in terms of their dual tensor multiplet models. 
The conformal theory (\ref{confhyper})  turns out to be dual to the improved $\cN=2$ tensor
model  \cite{KLR,LR2,deWPV,GIO}. When realized in the $\cN=2$ AdS superspace,
the latter is described by the following Lagrangian:
\be
\cL^{++}_{\rm impr.-tensor} =- G^{++} \ln \frac{G^{++}}{{\bm S}^{++} }~,
\ee
with $G^{++} $ a real $O(2)$ multiplet. 
The non-conformal theory  (\ref{L-0-massless}) is dual to the tensor multiplet 
model 
\be
\cL^{++}_{\rm tensor} =-\hf  \frac{(G^{++})^2}{{\bm S}^{++} }~.
\label{non-conf-tensor}
\ee
This is similar to the situation in $\cN=1$ AdS supersymmetry, 
where the conformal scalar multiplet model described by the Lagrangian 
\be 
L_{\rm conf} = {\bar \F}\F 
\ee
is dual to the improved tensor multiplet model \cite{deWR}
\be
L_{\rm impr.-tensor} =- G \ln G ~, \qquad 
\(\CDB^2-4\mu\) G=0 ~, \qquad G=\bar G~,
  \ee
while the non-conformal scalar multiplet model
\be 
L_{\rm non-conf} = \hf \big( {\bar {\bm \F}}+{\bm \F} \big)^2 
\ee
is dual to the ordinary tensor multiplet model \cite{Siegel}
\be
L_{\rm tensor} =-\hf G^2 ~.
\ee

A nonlinear generalization of the tensor multiplet model (\ref{non-conf-tensor}) is 
\be
\cL^{++} = {\bm S}^{++} \, F\Big(   \frac{G^{++}}{{\bm S}^{++} }\Big)~,
\ee
for some  function $F$, compare with the rigid $\cN=2$ 
supersymmetric models for tensor multiplets \cite{KLR}.
This theory can be seen to be dual to a weight-zero polar multiplet model of the form 
\be
\cL^{++} = {\bm S}^{++} \, {\mathbb F}\Big(  \widetilde{\bm\U} + {\bm\U}\Big)~,
\ee
for some  function $\mathbb F$ related to $F$.

\subsection{Models involving the intrinsic vector multiplet}

The structure of off-shell vector multiplets in a background of $\cN=2$ conformal supergravity
is discussed in \cite{KLRT-M}; see also 
Appendix B.
In the case of  AdS$^{4|8}$, we have $\cS^{ij}={\bar \cS}^{ij}={\bm S}^{ij}$.
Then, the Bianchi identity for the field strength $\cW$ of an Abelian vector multiplet,  
eq. (\ref{vectromul}), 
tells us that there exists 
a vector multiplet with a constant field strength,  $\cW_0$, which 
can be  chosen to be
\be
\cW_0=1~.
\ee 
Its existence is supported by the geometry of the AdS superspace, and for this reason 
 this vector multiplet will be called intrinsic. 
We denote  the corresponding tropical prepotential   by $V_0(z,u^+)$, 
and it should be emphasized  that $V_0$ is defined modulo  
gauge transformations of the form: 
\be
\d V_0 =\l  + \widetilde{\l}~,
\label{gaugeinvV0}
\ee
where $\l$ is  a covariant weight-zero arctic multiplet. 
Using $V_0$ allows us to construct a number of interesting models
in AdS$^{4|8}$.

Consider a system of Abelian vector supermultiplets in AdS$^{4|8}$
described by their covariantly chiral field strengths $\cW_I$, where $I=1,\dots, n$.
The dynamics of this system can be described by a Lagrangian of the form:
\bea
\cL^{++}=-\frac{1}{4} V_0 \Big{[} \Big(  (\cD^+)^2+4\bmS^{++} \Big) \cF (\cW_I)
+\Big( (\cDB^+)^2+4\bmS^{++}\Big){\bar \cF} ( \bar{\cW}_I) \Big{]}
~,~~~
\label{special-vector}
\eea
with $ \cF (\cW_I)$ a holomorphic function.
The action generated by $\cL^{++}$ is invariant under the gauge transformations 
(\ref{gaugeinvV0}).
This theory  is an AdS extension of the famous vector multiplet model 
behind the concept of  rigid special geometry \cite{rsg}.
The Lagrangian (\ref{special-vector}) 
is analogous to the rigid harmonic superspace representation 
for  effective vector multiplet models  given in \cite{DKT}.
In section 5, we will return to a  study of the model (\ref{special-vector}) 
for the case when AdS${}^{4|8}$ is replaced by a
general conformally flat superspace.

To describe massive hypermultiplets, we can follow the 
construction originally developed in the $\cN=2$ super-Poincar\'e case 
within the harmonic superspace approach \cite{vev} and later generalized 
to the projective superspace \cite{projective3,K06}.
That is, off-shell hypermultiplets should simply be coupled to the intrinsic vector multiplet, 
following the general pattern of coupling polar hypermultiplets to vector multiplets 
\cite{LR}. 
A massive weight-one polar hypermultiplet can be described by the Lagrangian
 \bea
\cL^{++}_1={\rm i}\,
\widetilde{\U}^+\re^{m\,V_0}{\U}^+~, \qquad m ={\rm const}
 \label{massive-conformal- hyper}
\eea
which is invariant under the gauge transformation of $V_0$, eq. (\ref{gaugeinvV0}), 
accompanied by
\be
\d \U^+ = - {m}\, \l\, \U^+~.
\ee 
Similarly, a massive weight-zero polar   multiplet 
can be described by the gauge-invariant Lagrangian:
\bea
\cL^{++}_2=\frac{\bmS^{++}}{|\bmS |}
\widetilde{\bm \U}\re^{m\,V_0}{\bm \U}~, \qquad m ={\rm const}~.
 \label{massive-hyper}
\eea

\section{Conformal flatness and  intrinsic vector multiplet} 
\setcounter{equation}{0}

We have seen that the dynamics of various models in AdS$^{4|8}$
is formulated using the  prepotential of the intrinsic vector multiplet. 
To reduce such actions to the $\cN=1$ AdS superspace, it is advantageous 
to realize AdS$^{4|8}$ as a conformally flat superspace. 

The fact that the $\cN=2$ AdS superspace is locally conformal flat
has already been discussed in the literature \cite{BILS}.
This result  will be re-derived  in a more general setting
in subsection 4.1.

It is useful to start by recalling the structure of super-Weyl transformations 
in 4D $\cN=2$ conformal supergravity following \cite{KLRT-M}. 
The superspace geometry describing   the 4D $\cN=2$ Weyl 
multiplet was studied in detail in \cite{KLRT-M}, and a  summary is given in Appendix A.
The corresponding covariant derivatives $\cD_{\underline{A}}=(\cD_a,\cD_\a^i,{\bar \cD}^\ad_i)$ 
obey  the constraints (\ref{constr-1}), and the latter are solved
in terms  of the dimension 1 tensors $\cS^{ij},\,\cG_{\a\ad},\,\cY_{\a\b}$ and $\cW_{\a\b}$
and their complex conjugates, see eqs. (\ref{acr1}--\ref{acr3}).
Let $D_{\underline{A}}=(D_a,D_\a^i,\DB^\ad_i)$ be another set of covariant derivatives 
satisfying the same constraints (\ref{constr-1}), 
with $S^{ij},\,G_{\a\ad},\,Y_{\a\b}$ and $W_{\a\b}$ 
being the dimension 1 components of the torsion. 
The two supergeometries, which are associated with $\cD_A$ and $D_A$, 
are said to be conformally related (equivalently, they describe 
the same Weyl multiplet) if they are related by a 
super-Weyl transformation of the form:\footnote{In \cite{KLRT-M},
only the infinitesimal  super-Weyl transformation was given.}\begin{subequations}\bea\cD_\a^i&=&\re^{\hf\sba}\Big(D_\a^i+(D^{\g i}\s)M_{\g\a}-(D_{\a k}\s)J^{ki}\Big)~,
\label{Finite-D}\\
\cDB_{\ad i}&=&\re^{\hf\s}\Big(\DB_{\ad i}+(\DB^{\gd}_{i}\sba)\bar{M}_{\gd\ad}
+(\DB_{\ad}^{k}\sba)J_{ki}\Big)~,
\label{Finite-Db} 
\\
\cD_a
&=&\re^{\hf(\s+\sba)}\Big(
D_a
+{\ri\over 4}(\s_a)^{\a}{}_\bd(\DB^{\bd}_{k}\sba)D_\a^k
+{\ri\over 4}(\s_a)^{\a}{}_\bd(D_\a^k\s)\DB^\bd_{k}
-{1\over 2}\big(D^b(\s+\sba)\big)M_{ab}
\non\\
&&~~~~~~~~~~
+{\ri\over 8}(\ts_a)^{\ad\a}(D^{\b k}\s)(\DB_{\ad k}\sba)M_{\a\b}
+{\ri\over 8}(\ts_a)^{\ad\a}(\DB^{\bd}_{k}\sba)(D_\a^k\s)\bar{M}_{\ad\bd}
\non\\
&&~~~~~~~~~~
-{\ri\over 4}(\ts_a)^{\ad\a}(D_{\a}^k\s)(\DB_\ad^{l}\sba)J_{kl}
\Big)
~,~~~~~~~~
\label{Finite-D_c}
\eea
\end{subequations}
where the parameter  $\s$ is covariantly chiral $\DB^\ad_i\s=0$.
The  dimension-1 components of the torsion are related as follows:
\begin{subequations}
\bea
\cS_{ij}&=&\re^{\sba}\Big(S_{ij}
-{1\over 4}(D^\g_{(i}D_{\g j)}\s)
+{1\over 4}(D^\g_{(i}\s)(D_{\g j)}\s)\Big)
\label{Finite-S}~,\\
\cG_\a{}^\bd&=&
\re^{\hf(\s+\sba)}\Big(G_\a{}^\bd
-{\ri\over 4}(\s^c)_\a{}^\bd D_c(\s-\sba)
-{1\over 8}(D_\a^k\s)(\DB^{\bd}_k\sba)
\Big)
~,
\label{Finite-G}
\\
\cY_{\a\b}&=&\re^{\sba}\Big(Y_{\a\b}
-{1\over 4}(D^k_{(\a}D_{\b)k}\s)
-{1\over 4}(D^k_{(\a}\s)(D_{\b)k}\s)\Big)
\label{Finite-Y}~,\\
\cW_{\a\b}&=&\re^{\s}{W}_{\a \b}~.
\label{Finite-W}
\eea
\end{subequations}

The geometry $\cD_{\underline{A}}$ will be called {\it conformally flat} if 
the covariant derivatives $D_{\underline{A}}$ correspond to a flat superspace.

Consider a vector multiplet. With respect to 
the conformally related covariant derivatives 
$\cD_{\underline{A}}$ and $D_{\underline{A}}$,
it is characterized by different covariantly chiral field strengths
$\cW$ and $W$ obeying the equations:
\bea
{\bar \cD}^i_\ad \cW=0~, \qquad 
\big(\cD^{\g(i}\cD_\g^{j)}+4\cS^{ij}\big)\cW
&=& \big(\cDB_\gd^{(i}\cDB^{ j) \gd}+ 4\bar{\cS}^{ij}\big)\bar{\cW}~, 
\non \\ 
{\bar D}^i_\ad W=0~, \qquad
\big(D^{\g(i}D_\g^{j)}+4S^{ij}\big)W
&=& \big({\bar D}_\gd^{(i}{\bar D}^{ j) \gd}+ 4\bar{S}^{ij}\big)\bar{W}~.
\non
\eea
The field strengths are related to each other as follows \cite{KLRT-M}:
\be
\cW = {\rm e}^{\s}\,W~.
\label{W-super-Weyl}
\ee

Consider a covariant weight-$n$ projective  supermultiplet.
With respect to the conformally related covariant derivatives 
$\cD_{\underline{A}}$ and $D_{\underline{A}}$,
it is described by superfields $\cQ^{(n)}$ and $Q^{(n)}$ 
obeying the constraints
\be
\cD^+_{\a} \cQ^{(n)}  = {\bar \cD}^+_{\ad} \cQ^{(n)}  =0~, 
\qquad 
D^+_{\a} Q^{(n)}  = {\bar D}^+_{\ad} Q^{(n)}  =0~.
\ee
In the case of matter multiplets,  
these superfields  are related to each other  as follows\footnote{The super-Weyl 
transformation laws  (\ref{W-super-Weyl}) and (\ref{Q-super-Weyl}) have natural 
counterparts in the case of 5D $\cN=1$ supergravity \cite{KT-Msugra3}.}
\cite{KLRT-M}: 
\be
\cQ^{(n)} = {\rm e}^{\frac{n}{2}(\s +\bar \s) }\,Q^{(n)}~.
\label{Q-super-Weyl}
\ee

As argued in \cite{KLRT-M}, the super-Weyl gauge freedom can always be used 
to impose the reality condition $\cS_{ij} = {\bar \cS}_{ij}$. 
The same condition can be chosen for the supergeometry generated by 
the covariant derivatives $D_{\underline{A}}$.
Therefore, if the conformally related supergeometries are characterized by 
the reality conditions
\be
\cS_{ij} = {\bar \cS}_{ij}~, \qquad S_{ij} = {\bar S}_{ij}~,
\ee
then eq. (\ref{Finite-S}) tells us that 
\be
W:= {\rm e}^{-\s} 
\ee
is the covariantly chiral field strength of a vector multiplet with respect 
to the covariant derivatives $D_{\underline{A}}$.
Due to (\ref{W-super-Weyl}), we then have $\cW=1$.

It is instructive to compare the $\cN=2$ super-Weyl transformation, eqs. 
(\ref{Finite-D}--\ref{Finite-D_c}), with that in $\cN=1$ 
old minimal supergavity \cite{HT}:
\begin{subequations}
\bea
\CD_\a&=&F D_\a-2(D^{\g}F)M_{\g\a}~, \qquad
F:=\vf^{1/2}\vfb^{-1}~, \qquad \DB^\ad \vf=0~
\label{N=1der-al}\\
\CDB_\ad&=&\bar{F}\DB_{\ad}-2(\DB^{\gd}\bar{F})\bar{M}_{\gd\ad}~,\\
\CD_{\a\ad}&=&{\ri\over 2}\{\CD_\a,\CDB_\ad\}~.
\label{N=1der-a}
\eea
\end{subequations}
Here $\nabla_A = (\nabla_a, \CD_\a,\CDB^\ad)$ and 
$D_A = (D_a, D_\a,{\bar D}^\ad)$ are two sets of  
$\cN=1$ supergravity covariant derivatives obeying the 
modified Wess-Zumino constraints.

\subsection{Reconstructing the intrinsic vector multiplet}

The superspace geometry of  AdS$^{4|8}$ is determined by 
the relations (\ref{AdS-geometry1}) and (\ref{AdS-geometry2}).
Let us demonstrate that  AdS$^{4|8}$ is conformally flat, 
which we note would imply that  
$S_{ij}=G_{\a\ad}=Y_{\a\b}=W_{\a\b}=0$
in eqs. (\ref{Finite-S}--\ref{Finite-W}).
Our first task is to 
search for a chiral scalar $\s$ such that $\cY_{\a\b}
=\cG_{\a\bd}=0$.
The equation $\cY_{\a\b}=0$ 
is equivalent to
\bea
D^k_{(\a}D_{\b)k}\,\re^{\s}=0~.~~~~~~
\label{Y=0}
\eea
The equation $\cG_{\a\bd}=0$ is equivalent to
\bea
{[}D_\a^k,\DB^\ad_k{]}\re^{\s+\sba}=0~.
\label{G=0}
\eea
The covariant derivatives of the flat global $\cN=2$ superspace are
$D_A =(\pa_a,D_\a^i,\DB^\ad_i)$, with 
\bea
D_\a^i&=&{\pa\over\pa\q_i^\a}-\ri(\s^b)_{\a}{}^{\bd}\qb_\bd^{i}\pa_b~, \qquad
\DB^{\ad}_{ i}={\pa\over\pa\qb_{\ad}^{i}}-\ri(\s^b)_\b{}^{\ad}\q^\b_{i}\pa_b~.
\eea

Consider a Lorentz invariant  ansatz for $\s$ and $\sba$ given by 
\bea
\re^\s=A(x_{\rm L}^2)+\q_{ij}B^{ij}(x_{\rm L}^2)+\q^4C(x_{\rm L}^2)~,\qquad 
\re^\sba=\bar{A}(x_{\rm R}^2)+\qb^{ij}\bar{B}_{ij}(x_{\rm R}^2)
+\qb^4\bar{C}(x_{\rm R}^2)~,
\label{ansatz}
\eea
where
\begin{subequations}
\bea
&x_{\rm L}^a:=x^a-\ri(\s^a)_{\a}{}^{\ad}\q^{\a}_k\qb_{\ad}^k~,\qquad
\q_{ij}:=\q^\a_i\q_{\a j}~,\qquad\q^4:=\q_{ij}\q^{ij}~,
\\
&x_{\rm R}^a:=x^a+\ri(\s^a)_{\a}{}^{\ad}\q^{\a}_k\qb_{\ad}^k~,\qquad
\qb^{ij}:=\qb_\ad^i\qb^{\ad j}~,\qquad \qb^4:=\qb^{ij}\qb_{ij}~,
\eea
\end{subequations}
and the functions $\bar{A},\,\bar{B}_{ij},\,\bar{C}$ are the complex conjugates of 
${A},\,{B}^{ij},\,C$.
The variables $x_{\rm L}^a$ and $\q^\a_i$ parametrize the chiral subspace of the
flat $\cN=2$ superspace. 

Equation (\ref{Y=0}) proves to restrict the coefficients in (\ref{ansatz}) to look like 
\bea
A(x^2)=a_1+a_2\,x^2~,\qquad B^{ij}(x^2)=b^{ij}~,\qquad C(x^2)=0~,
\label{Y=0-sol}
\eea
where $a_1,\,a_2,\,b^{ij}$ are constant parameters.
Next, equation (\ref{G=0}) imposes additional conditions on the parameters $a_1,\,a_2$
and $b^{ij}$:
\bea
a_1\bar{a}_2=\bar{a}_1a_2~,\qquad
b^{ik}\bar{b}_{kj}=-4a_1\bar{a}_2\d^i_j~.
\label{G=0-sol}
\eea
Without loss of generality, 
the constant $a_1$ can be chosen to be 
$a_1=1$, 
and then the relations (\ref{G=0-sol}) are equivalent to
\bea
b^{ij}=q \bms^{ij}~, \quad a_2=-{1\over 4}\bms^2~, \qquad 
\overline{\bms^{ij}}=\bms_{ij}~,\quad |q|=1~,\quad
\bms^2:=\hf\bms^{ij}\bms_{ij}~.
~~~~~
\eea
It can be seen that  the parameter $q$ coincides with that appearing 
in (\ref{AdS-geometry2}). In accordance with the consideration in section 2, 
we set $q=1$.
Now, the solution to eqs. (\ref{Y=0}) and (\ref{G=0}) can be expressed as 
\bea
\re^\s=1-{1\over4}\bms^2x_{\rm L}^2+ \bms^{ij}\q_{ij}~,\qquad
\re^\sba=1-{1\over4}\bms^2x_{\rm R}^2+ \bms_{ij}\qb^{ij}~.
\label{Solution-AdS}
\eea

Note that the tensors $\cS^{ij}$ and $\bar{\cS}^{ij}$ are expressed in terms of $\s$ 
and $\bar \s$ as follows:
\bea
\cS^{ij}&=&{1\over 4}\re^{\s+\sba}(D^{ij}\re^{-\s})~,\qquad
\bar{\cS}^{ij}={1\over 4}\re^{\s+\sba}(\DB^{ij}\re^{-\sba})
~,
\label{S^{ij}-0}
\eea
with
$D^{ij}:=D^{\g(i}D_\g^{j)}$ and 
$\DB^{ij}:=\DB_\gd^{(i}\DB^{j)\gd}$.
It also holds
\bea
\cS^{ij}&=& \bms^{ij}~+~O(\q)~,\qquad
\bar{\cS}^{ij}= \bms^{ij}~+~O(\q)
~.
\eea
Then, the relation 
\be
\cS^{ij}=\bar{\cS}^{ij} \equiv {\bm S}^{ij}
\label{SSS}
\ee
holds as a consequence of the Bianchi identities.
Defining a new 
chiral superfield
\bea
W_0:= \re^{-\s} = \Big(1-{1\over4}\bms^2x_{\rm L}^2+ \bms^{ij}\q_{ij}\Big)^{-1}
~,
\qquad 
\DB^\ad_iW_0
=0~,
\label{Solution-AdS2}
\eea
one can see that eq. (\ref{SSS}) is equivalent to 
\bea
D^{ij}W_0=\DB^{ij}\bar{W}_0~.
\label{W_0-vector}
\eea
This is the Bianchi identity for the field strength of 
an Abelian vector multiplet 
in flat superspace \cite{GSW}.
It is an instructive exercise to check 
eq. (\ref{W_0-vector}) 
by explicit calculations. 

It follows from the expression for $W_0$, eq. (\ref{Solution-AdS2}),
and the explicit form for the vector covariant derivative $\cD_a$, 
eq. (\ref{Finite-D_c}), that the space-time metric is
\be
{\rm d}s^2 = {\rm d}x^a  \,{\rm d}x_a \,\Big(W_0{\bar W}_0\Big) \Big|_{\q=0}
= \frac{{\rm d}x^a  {\rm d}\,x_a }{\big(1-{1\over4}\bms^2x^2\big)^2}~.
\label{metric}
\ee  
Modulo a trivial redefinition, this expression coincides 
with the metric in the north chart of AdS${}_4$ defined in Appendix D, 
with $x^a$ being the stereographic coordinates.
The metric can be brought to the form (\ref{metric-north})
by re-scaling $x^a \to 2x^a$ and then identifying $\bms^2 =R^{-2}$.
As expected,  the conformally flat representation 
(\ref{Finite-D}--\ref{Finite-D_c}) is defined only locally. 

Associated with the field strengths $W_0$ and $\bar{W}_0$
is their descendant 
\bea
\S_0^{ij}:=\frac{1}{ 4}D^{ij}W_0=\frac{1}{4}\DB^{ij}\bar{W}_0~, \qquad 
\overline{\S_0^{ij}} = \ve_{ik} \ve_{jl}\, \S_{0}^{kl}
\eea
enjoying the properties
\bea
D_\a^{(i}\S_0^{jk)}=\DB^{\ad(i}\S_0^{jk)}=0
\eea
that are  characteristic of the $\cN=2$ tensor multiplet.
Contracting the indices of $\S_0^{ij}$ with the isotwistor variables 
$u^+_i \in {\mathbb C}^2 \setminus \{0\}$, we then obtain the following real $O(2)$ multiplet: 
\bea
\S_0^{++}(z,u^+):=u_i^+u_j^+\S_0^{ij}(z)~,
\qquad D_\a^+\S_0^{++}=\DB^{\ad+}\S_0^{++}=0~.
\label{Sigma_0-0}
\eea
It can be shown that 
$\S_0^{++}$ has the form:
\bea
\S_0^{++}&=&
{\bms^{++}\over \Big(1-{1\over 4}\bms^2x_{\rm A}^2\Big)^2}
-{2\bms^2\big((\q^{+})^2+ (\qb^{+})^2\big)\over \Big(1-{1\over 4}\bms^2x_{\rm A}^2\Big)^3}
-{2\ri \bms^2  \bms^{+-} (x_{\rm A} )_\a{}^{\ad}\q^{\a+}\qb_{\ad}^{+}
\over (u^+u^-)\Big(1-{1\over 4}\bms^2x_{\rm A}^2\Big)^3}
\non\\
&&
+\hf \frac{\bms^2 \bms^{--}\big(8+ \bms^2 x_{\rm A}^2 \big)(\q^{+})^2 (\qb^{+})^2
 }{(u^+u^-)^2\Big(1-{1\over 4}\bms^2x_{\rm A}^2\Big)^4}
~.
\label{Sigma_0^{++}-analytic}
\eea
Here ${\bm s}^{\pm \pm }= {\bm s}^{ij} u^\pm_i u^\pm_j$, 
$\q^\pm_\a = \q^i_\a u^\pm_i$ and ${\bar \q}^\pm_{\ad} = {\bar \q}^i_{\ad} u^\pm_i$,
$(\q^+)^2 = \q^{+\a}\q^+_\a$ and 
\bea
x_{\rm A}^a=x^a+{\ri\over(u^+u^-)}(\s^a)_{\a}{}^\ad \Big(\q^{\a+}\qb_\ad^-
+
\q^{\a-}\qb_\ad^+ \Big)~.
\eea
The variables $x_{\rm A}^a$, $\q^+_\a$ and ${\bar \q}^+_{\ad}$ 
are annihilated by the covariant derivateves
$D^+_\a:=u_i^+D_\a^i$ and $\DB^{\ad+}:=u_i^+\DB^{\ad i}$, and 
can be used to parametrize the analytic subspace 
of harmonic superspace \cite{GIKOS,GIOS}.
One can check that $\S^{++}_0$ has the form (\ref{Sigma_0-0}), and hence 
does not depend on $u^-$, 
\be
\frac{\pa }{\pa u^-} \S_0^{++}=0~, 
\ee
in spite of the fact that separate contributions
to the right-hand side of (\ref{Sigma_0^{++}-analytic}) explicitly depend
on $u^-$.

In conclusion, we give the explicit expression for the torsion ${\bm S}^{ij}$:
\bea
{\bm S}^{ij}=(W_0 \bar{W}_0)^{-1}\, \S_0^{ij}~.
\eea
It is important to point out that  now ${\bm S}^{ij}$ is  covariantly 
constant, $\cD^i_\a {\bmS}^{kl} = {\bar \cD}^i_\ad {\bmS}^{kl}=0$,  
but not constant. This clearly differs from the analysis in section 2,  
and the origin of this disparity  is very simple. 
In section 2, we imposed the SU(2) gauge (\ref{SU(2)conn})
in which only a U(1) part of the SU(2) connection survived, 
and the covariant derivatives had the form (\ref{cov-der-U(1)-rep}).
Here we are using the conformally flat representation 
for the covariant derivatives, eqs. (\ref{Finite-D})  and (\ref{Finite-Db}), 
such that the connection becomes a linear combination of all the generators 
of the group SU(2).

\subsection{Prepotential for the intrinsic vector multiplet}

The field strength $W_0$ of the intrinsic vector 
multiplet, eqs. (\ref{Solution-AdS}) and (\ref{Solution-AdS2}), 
depends on the {\it constant} isotensor ${\bm s}^{ij} =  {\bm s}^{ji}$
obeying the reality condition $\overline{{\bm s}^{ij}} = {\bm s}_{ij}$.
By applying a rigid SU(2) rotation one can always set
\bea
\bms^{\1\2}=0~.
\label{s12}
\eea
This choice will be  used in the remainder of the paper.

Modulo gauge transformations, the prepotential for the intrinsic vector 
multiplet can be chosen to be 
\bea
V_0(z,u^+)=
V_0(z,\z)&=&\ri {{\bm\q}^2(\z)+{\bm\qb}^2(\z)\over \z 
\Big(1-{1\over 4}|\bms^{\1\1}|^2x_{\rm  A}^2(\z)\Big)}
-{\rm i}{\big(\z\bms^{\1\1}+{1\over \z}
\bms^{\2\2}\big){\bm\q}^2(\z){\bm\qb}^2(\z)\over  \z^2
\Big(1-{1\over 4}|\bms^{\1\1}|^2x_{\rm A}^2(\z)\Big)^2}
~.~~~
\label{V_0-proj}
\eea
Here we have made use of  the 
complex coordinate $\z$ for  ${\mathbb C}P^1$ as well as the following
$\z$-dependent superspace variables
\bea
&{\bm\q}^{\a}(\z)=-\z\q^{\a}_\2-\q^\a_\1~,\qquad
{\bm\qb}_{\ad}(\z)=-\z\qb_{\ad}^{\1}+\qb_{\ad}^{\2}~, \non \\
&x_{\rm A}^a(\z)=x^a+\ri(\s^a)_{\a}{}^\ad{\bm\q}^{\a}(\z)\qb_\ad^\1
+\ri(\s^a)_{\a}{}^\ad\q^{\a}_\2{\bm\qb}_\ad(\z)~,
\eea
which are annihilated by  $\z_iD_\a^i$ and $\z_i\DB^{\ad i}$,
with $\z_i=(-\z,1)$.

\subsection{$\cN=1$ reduction revisited}

We have elaborated upon the superspace reduction $\cN=2 \to \cN=1$ 
in subsection 2.2 using the representation (\ref{cov-der-U(1)-rep})
for  the covariant derivatives.
Such a reduction should be carried out afresh 
if the covariant derivatives are given 
in the conformally flat representation 
defined by  eqs. (\ref{Finite-D})  and (\ref{Finite-Db}). 
One of the reasons for this is that the component $\bmS^{\1\2}$
of the torsion $\bmS^{ij}$ does not vanish 
and the algebra of the operators
$(\cD_a,\cD_\a^\1,\cDB^\ad_\1)$
is no longer closed, for the third relation in (\ref{D^1-alg}) turns into
\bea
{[}\cD_a,\cD_\b^\1{]}&=&
{\ri\over 2}({\s}_a)_{\b\gd}\bmS^{\1\1}\cDB^\gd_\1
+{\ri\over 2}({\s}_a)_{\b\gd}\bmS^{\1\2}\cDB^\gd_\2~.
\label{reduction-AdS-0}
\eea
Nevertheless, it can be shown, using (\ref{s12}), that
the projection of $\bmS^{\1\2}$ does vanish, 
\bea
\bmS^{\1\2}\big|=0~.
\label{S^{12}-reduced}
\eea
Another consequence of  the choice (\ref{s12}) is 
\bea
(D_{\a}^\2\s)|=(\DB^{\ad}_{\2}\sba)|=0~.
\eea
Then, applying the $\cN=1$ projection to  the covariant derivatives, 
\bea
\cD_{\underline{A}}|:=\cE_{\underline{A}}{}^{\underline{M}}|\pa_{\underline{M}}]
+\hf \O_{\underline{A}}{}^{bc}|M_{bc}+\F_{\underline{A}}{}^{kl}|J_{kl}~,
\eea
for $\cD_\a^\1|$ and $\cDB_{\ad \1}|$ we get
\begin{subequations}
\bea
\cD_\a^\1|&=&\re^{\hf\sba|}\Big(D_\a+(D^{\g}\s|)M_{\g\a}+(D_{\a}\s|)J_{\1\2}\Big)~,
\label{reduction-1}
\\
\cDB_{\ad \1}|&=&\re^{\hf\s|}\Big(\DB_{\ad}+(\DB^{\gd}\sba|)\bar{M}_{\gd\ad}
+(\DB_{\ad}\sba|)J_{\1\2}\Big)~.
\label{reduction-2}
\eea
\end{subequations}
Here $D_\a$ and ${\bar D}^\ad$ are the spinor covariant derivatives
for the flat global $\cN=1$ superspace parametrized by 
$(x^a,\q^\a,\qb_\ad)$, with 
\bea
\q^\a:=\q^\a_\1~, \quad \qb_\ad:=\qb_\ad^\1~, \qquad
D_\a:=D_\a^\1|~,  \quad \DB^\ad:=\DB^\ad_\1|~.
\eea
As is seen from (\ref{reduction-1}) and (\ref{reduction-2}), 
the operators $\cD_\a^\1|$ and $\cDB_{\ad \1}|$ do not involve any 
partial derivatives with respect to $\q_\2$ and ${\bar \q}^\2$. 
Another important property is that  the operator $J_{\1\2}$ is diagonal when acting on 
$\cD_\a^\1$ and $\cDB_{\ad \1}$.
Therefore, for any positive integer $k$,  
it holds that $\big( \cD_{\hat{\a}_1} \cdots  \cD_{\hat{\a}_k} U \big)\big|
= \cD_{\hat{\a}_1}| \cdots  \cD_{\hat{\a}_k}| U|$, 
where $ \cD_{\hat{\a}} =( \cD_\a^\1, \cDB^\ad_\1)$ and $U$ is an arbitrary superfield.
This implies that the operators  $(\cD_a|,\cD_\a^\1|,\cDB^\ad_\1|)$ 
satisfy the (anti-)commutation relations:
\bea
\big\{\cD_\a^\1|,\cD_\b^\1| \big\}&=&
4{\bmS}^{\1\1}|M_{\a\b}~,\qquad
\big\{\cD_\a^\1|,\cDB^\bd_\1| \big\}=
-2\ri(\s^c)_\a{}^\bd\cD_c|
~,
\non\\
\big{[}\cD_a|,\cD_\b^\1|\big{]}&=&
{\ri\over 2}({\s}_a)_{\b\gd}\bmS^{\1\1}|\cDB^\gd_\1|~.
\label{reduction-AdS}
\eea

The algebra (\ref{reduction-AdS}) is isomorphic to that of the $\cN=1$ 
AdS covariant derivatives $\nabla_A =(\nabla_a,\nabla_\a,{\bar \nabla}^\ad)$, 
see Appendix C. Unlike $\nabla_A$, however,
 the operators  $(\cD_a|,\cD_\a^\1|,\cDB^\ad_\1|)$ involve a zero-curvature U(1) connection,
 with $J_{\1\2}$ the U(1) generator. The latter connection can be gauged away.
Making use of the explicit action of the generator $J_{\1\2}$ 
on the covariant derivatives, 
\bea
[J_{\1\2},\cD^\1_\a]=-\hf \cD_\a^\1~, \qquad
[J_{\1\2},\cDB^\ad_{\1}]=\hf\cDB^\ad_{\1}~.
\eea
one finds
\bea
\re^{-(\sba-\s)|J_{\1\2}} \, \cD_\a^\1| \,\re^{(\sba-\s)|J_{\1\2}}
=\CD_\a
~,\qquad
\re^{-(\sba-\s)|J_{\1\2}} \,
\cDB_{\ad \1}| \, \re^{(\sba-\s)|J_{\1\2}}=\CDB_\ad~.
\eea
Here the operators $\nabla_A = (\nabla_a, \CD_\a,\CDB^\ad)$ 
have the form (\ref{N=1der-al}--\ref{N=1der-a}), where 
$D_A = (D_a, D_\a,{\bar D}^\ad)$ are the flat $\cN=1$ covariant derivatives, 
and the chiral superfield $\vf$ is
\bea
\vf:=W_0|=\Big(1-{\mu\bar{\mu}\over4}x_{\rm L}^2-\mub\q^2\Big)^{-1}~,\qquad
\DB^\ad\vf=0~,
\label{vf}
\eea
with
\bea
\mub:=-\bms^{\1\1}~, \qquad
\mu:=-{\bms}^{\2\2}=-\bms_{\1\1}~.
\label{s-sb}
\eea
The operators $\nabla_A = (\nabla_a, \CD_\a,\CDB^\ad)$
coincide with the $\cN=1$ AdS covariant derivatives
as given in \cite{BK}, and   satisfy 
the (anti-)commutation relations 
(\ref{N=1-AdS-algebra-1}) and (\ref{N=1-AdS-algebra-2}).

Let us describe
the action of the U(1)-rotation $\re^{-(\sba-\s)J_{\1\2}}$ 
on different types of projective multiplets.
For a covariant  weight-$n$ arctic hypermultiplet  (\ref{arctic-n})
it holds
\bea
\U^{[n]}(z,\z)=\sum_{k=0}^{+\infty}\U_k(z)\z^k~,\qquad
\re^{-(\sba-\s) J_{\1\2}}\U^{[n]}(z,\z)=\re^{-\frac{n}{2}(\sba-\s)}\U^{[n]}
\big(z,\re^{(\sba-\s)}\z\big)~.~~~
\eea
Here we have used the results of \cite{KLRT-M} for the SU(2)-transformation 
rules of the component superfields of projective multiplets.
In the case of a real weight-2n projective superfield (\ref{real-2n}), such as $O(2n)$ multiplets,
one finds
\bea
R^{[2n]}(z,\z)=\sum_{k=-\infty}^{+\infty}R_k(z)\z^k~, \qquad
\re^{-(\sba-\s)J_{\1\2}}R^{[2n]}(z,\z)=R^{[2n]}\big(z,\re^{(\sba-\s)}\z\big)~.~~~
\label{J_12-R-2n}
\eea

To conclude this section, we wish to  give  the expressions for
$\bmS^{ij}|$ and $V_0|$ which will be useful in what  follows.
For the $O(2)$ multiplet $\bmS^{++}:=u^+_iu^+_j\bmS^{ij}$,
one can show
\bea
\bmS^{++}|=\ri u^{+\1}u^{+\2}\bmS(\z)|~, \qquad
\bmS(\z)|=\ri\Big(\vf^{-1}\vfb\,\mub\,\z+\vfb^{-1}\vf\,\mu\,{1\over\z}\Big)~.
\eea
It is important to note that
\bea
\re^{-(\sba-\s)|J_{\1\2}}\bmS(\z)|=\ri\(\mub\,\z+\mu\,{1\over\z}\)~,
\eea
where we have used (\ref{J_12-R-2n}).
${}$For the prepotential $V_0(\z)$ of the intrinsic vector multiplet,
we obtain
\bea
V_0(\z)|&=&\ri\(\vf\qb^2\,\z+\vfb \q^2\,{1\over \z}\)~,
\eea
and hence
\bea
\hat{V}_0(\z):=\re^{-(\sba-\s)J_{\1\2}}V_0(\z)|&=&\ri\(\big({\vf^2\vfb^{-1}}\qb^2\big)\z
+\big({\vfb^2\vf^{-1}}\q^2\big){1\over \z}\):=
\z V_+
-{1\over\z}V_-
~.
\label{V_0-reduced}
\eea

\subsection{$\cN=2$ AdS Killing supervectors: II}

In this subsection, the  $\cN=2$ AdS Killing supervectors
are explicitly evaluated using the conformally flat
representation for $\cD_{\underline{A}}$ derived earlier.

Our starting point will be the observation that
the conformally related supergeometries have isomorphic
superconformal algebras (see \cite{BK} for
a pedagogical discussion of this result in the case of 4D $\cN=1$
supergravity). Therefore, since the superspaces  ${\mathbb R}^{4|8}$
and AdS${}^{4|8}$ are conformally related, they possess the same superconformal
algebra, su$(2,2|2)$. It is well known how su$(2,2|2)$ is realized in the 4D $\cN=2$
flat superspace, see e.g. \cite{K-hyper1,K-hyper2,Ferber,Park,KT} and references therein.
Let us first recall this realization following \cite{K-hyper1,K-hyper2,KT}.

By definition, a superconformal  Killing vector of ${\mathbb R}^{4|8}$
\be
{\bm \x} ={\bar {\bm \x} }={\bm \x}^{\underline{A}} (z) D_{\underline{A}}
= {\bm \x}^a  \pa_a + {\bm \x}^\a_i D^i_\a
+ {\bar {\bm \x}}_{\dot \a}^i  {\bar D}^{\dot \a}_i
\ee
obeys the constraint
\bea
\d_{\bm \s} D_{\underline{A}}
+ [ {\bm \x} +\hf K^{cd}M_{cd} +K^{kl}J_{kl} , D_{\underline{A}}\big] =0~,
\eea
for a chiral scalar ${\bm \s} (z)$, ${\bar D}^\ad_i {\bm \s}=0$,
which  generates an infinitesimal super-Weyl transformation,
a real antisymmetric tensor $K^{cd}(z)$ and
a real symmetric tensor $K^{kl}(z)$.
This constraint  implies
\bea
{\bar D}^\ad_i K^{\b \g}&=&
0~,  \qquad
D^i_\a K^{\b \g} = \d_\a^{(\b} D^{\g) i} {\bm \s} ~,\qquad
D^i_\a K^{kl} = \ve^{i (k} D^{l) }_\a {\bm \s} ~,
\eea
as well as
\be
[{\bm \x} ,D^i_\a ]
=  -\hf \bar{ {\bm \s}} \, D^i_\a
-K_\a{}^\b  D^i_\b
- K^i{}_j D^j_\a ~.
\label{4DmasterN=2}
\ee
The latter equation, in turn, leads to
\begin{subequations}
\bea
K_{\a \b} &=& \frac{1}{2}\;D^i_{(\a} {\bm \x}_{\b)i}~,
\qquad {\bm \s}  = \frac{1}{2}
{\bar D}^{\dot \a}_i {\bar {\bm \x}}_{\dot \a}^{ i} ~,\\
K^i{}_j &=& \hf \Big(D^i_\a {\bm\x}^\a_j - \hf \d^i_j D^k_\a {\bm\x}^\a_k \Big)
=-\hf \Big({\bar D}^{\dot \a}_j {\bar {\bm\x}}^i_{\dot \a}
-\hf \d^i_j {\bar D}^{\dot \a}_k {\bar {\bm\x}}^k_{\dot \a} \Big)~,
\eea
\end{subequations}
as well as
\be
{\bar D}^\ad_i {\bm \x}^{\bd \b} = 4{\rm i} \ve^{\ad \bd} {\bm \x}^\b_i~, \qquad
{\bar D}^\ad_i {\bm \x}^\b_j=0~.
\ee
The general expression for the superconformal Killing vector can be shown to be
\bea
{\bm\x}^a&=&\hf\({\bm\x}_{\rm L}^a+\bar{\bm \x}_{\rm R}^a\)
+\ri(\s^a)_\a{}^\ad{\bm\x}^\a_k\qb^k_\ad
+\ri(\s^a)_\a{}^\ad\bar{\bm\x}_\ad^k\q_k^\a
~,  \non\\
{\bm\x}_{\rm L}^{\ad\a}&=&p^{\ad\a}
+(r+\bar{r})x_{\rm L}^{\ad\a}
-\bar{\o}^\ad{}_\bd x_{\rm L}^{\bd\a}
-x_{\rm L}^{\ad\b}\o_\b{}^\a
+x_{\rm L}^{\ad\b}k_{\b\bd}x_{\rm L}^{\bd\a}
+4\ri\bar{\e}^{\ad k}\q^\a_k
-4x_{\rm L}^{\ad\b}\eta_\b^k\q^\a_k~,~~~~~\non \\
{\bm\x}^\a_i&=&
\e^\a_i
+\bar{r}\q^\a_i
-\q^\b_i\o_\b{}^\a
-\L_i{}^j\q^\a_j
+\q^\b_i k_{\b\bd}x_{\rm L}^{\bd\a}
-\ri\bar{\eta}_{i\bd}x_{\rm L}^{\bd\a}
-4\q^\b_i\eta_\b^k\q^\a_k~,
\label{conf-Killing}
\eea
see, e.g., \cite{Park,K-hyper1} for two different derivations.
Here the  constant parameters $(\o_\a{}^\b,~{\bar \o}^{\dot \a}{}_{\dot \b})$ correspond to a
Lorentz transformation,
$p^{\dot \a \b}$ a space-time translation,
$ k_{\a \dot \b}$ a special conformal transformation,
$r$ a combined scale and chiral U(1) transformation,
$(\e_i^\a,~ {\bar \e}^{\dot \a i})$ and
$(\eta_\a^i,~{\bar \eta}_{i \dot \a})$
$Q$--supersymmetry  and $S$--supersymmetry
transformations respectively,
and finally $\L_i{}^j$ an SU(2) transformation.

If $W$ is the chiral field strength of an Abelian vector multiplet
in ${\mathbb R}^{4|8}$, such that
$D^{\a i}D_\a^{j}W= {\bar D}_\ad^{i}{\bar D}^{ j \ad}\bar{W} $
is the corresponding Bianchi identity,
its superconformal transformation is
\be
\d W= {\bm \x} W + {\bm \s} W~,
\ee
see, e.g., \cite{KT}. The superconformal transformations
of the rigid projective multiplets are given in \cite{K-hyper2}.

Now, let us return to the $\cN=2$ AdS superspace, and let
$\x^{\underline{A}} (z) \cE_{\underline{A}}$ be its
Killing supervector. We can represent
\be
\x^{\underline{A}} (z) \cE_{\underline{A}}
= {\bm \x}^{\underline{A}} (z) D_{\underline{A}} \equiv {\bm \x}~,
\ee
where
\bea
{\bm\x}^a&=&\re^{\hf(\s+\sba)}\x^a~,
\qquad
{\bm\x}^\a_i=\re^{\hf\sba}\x^\a_i
+{\ri\over 4}\re^{\hf(\s+\sba)}\x^{\a}{}_\bd(\DB^{\bd}_{i}\sba)~.
\label{Killing-1-2}
\eea
Then, eq. (\ref{Super-K-eq-00}) proves to be equivalent  to the fact that $\bm \x$ is
a superconformal Killing supervector in ${\mathbb R}^{4|8}$ such that
\be
\d W_0= {\bm \x} W_0 + {\bm \s} W_0=0~,
\label{masterAdS}
\ee
with $W_0$ the field strength of the intrinsic vector multiplet.
In other words, $W_0$ is invariant under the $\cN=2$ AdS transformations
(which is completely natural, keeping in mind  that $\cW_0 =1$).
The invariance of $W_0$ implies that the AdS transformation of the
prepotential $V_0$ is a pure gauge transformation.

The general solution of (\ref{masterAdS}) can be shown to be
\begin{subequations}
\bea
r&=&0~,
\label{Killing-solved-1}
\\
k^a&=&\frac{1}{4}\bms^2p^a~,
\label{Killing-solved-2}
\\
\eta_\a^i&=&\hf\bms^{ij}\e_{\a j}~,~~~\bar{\eta}^\ad_i=\hf\bms_{ij}\bar{\e}^{\ad j}~,
\label{Killing-solved-3}
\\
\L_{ij}&=&l\bms_{ij}~,~~~~~~\bar{l}=l~,
\label{Killing-solved-4}
\eea
\end{subequations}
with no restrictions on the Lorentz parameters.
Using the solution (\ref{Killing-solved-1})--(\ref{Killing-solved-4}) in (\ref{conf-Killing}),
from (\ref{Killing-1-2}) one can read the $\cN=2$ AdS Killing supervectors $\x$ 
in terms of ${\bm\x}$.

It is instructive to 
consider the $\cN=1$ reduction 
of the $\cN=2$ AdS Killing supervectors.
Let us first give the $\cN=1$ projection of the superconformal 
Killing vector ${\bm\x}$ associated   with the $\cN=2$ 
AdS Killing vector field $\x^{\underline{A}} (z) \cE_{\underline{A}}$:
\begin{subequations}
\bea
{\bm\l}^{\a\ad}&=&{\bm\x}^{\a\ad}|=
\Big(1-{{|\mu|^2}\over 4}\q^2\qb^2\Big)p^{\a\ad}
+{|\mu|^2\over 4}x^{\a\bd}p_{\b\bd}x^{\b\ad}
-\o^\a{}_\b x^{\b\ad}
-\bar{\o}^\ad{}_\bd x^{\a\bd}
-2\q^\a(2\ri\bar{\e}^{\ad\1}
+\mub x^{\b\ad}\e_{\b\1})
\non\\
&&~~~~~~~~
-2\qb^\ad(2\ri\e^\a_{\1}
-\mu x^{\a\bd}\bar{\e}_\bd^{\1})
-\ri\q^\a\qb_\bd\Big(2\bar{\o}^{\ad\bd}
+{|\mu|^2\over 2}p^{\b(\ad} x_{\b}{}^{\bd)}\Big)
\non\\
&&~~~~~~~~
-\ri\qb^\ad\q_\b\Big(2{\o}^{\a\b}
-{|\mu|^2\over 2}p^{(\a}{}_\bd x^{\b)\bd}\Big)
-2\ri\mub\e^\a_{\1}\qb^\ad\q^2
-2\ri\mu\bar{\e}^{\ad\1}\q^\a\qb^2
~,~~~~~~
\\
{\bm\l}^\a&=&{\bm\x}^\a_\1|=
\e^\a_\1\big(1-\mub\q^2\big)
-\q^\b\o_\b{}^\a
+{|\mu|^2\over 4}\q^\b p_{\b\bd}x_L^{\bd\a}
+{\ri\over 2}\mu\bar{\e}_{\bd}^\1x_L^{\bd\a}
~,
\\
{\bm\ve}^\a&=&{\bm\x}^\a_\2|=
\e^\a_\2
+l\mub\q^\a
+{\ri\over 2}\mub\bar{\e}_{\bd}^\2x_L^{\bd\a}~.
\label{bolde}
\eea
\end{subequations}
Then, the $\cN=1$ AdS Killing supervector 
$\L=\l^a\CD_a+\l^\a\CD_\a+\bar{\l}_\ad\CDB^\ad$ 
is expressed in terms of ${\bm\l}_a$ and ${\bm\l}^\a$ 
as follows:
\begin{subequations}
\bea
\l^a&=&\vf^\hf\bar{\vf}^\hf{\bm\l}^a~,
\\
\l^\a&=&-{\ri\over 8}\CDB_\bd\l^{\a\bd}
=\vf^{-\hf}\bar{\vf}\Big({\bm\l}^\a
+{\ri\over 4}{\bm\l}^{\a}{}_\bd \DB^{\bd}\log{\bar{\vf}}\Big)~.
\eea
\end{subequations}
These expressions agree with \cite{BK}.
The second supersymmetry and U(1) transformations 
in the $\cN=1$ AdS superspace
are generated by $\ve $ and $\ve^\a$ 
which are related to ${\bm\ve}^\a$ 
appearing in eq. (\ref{bolde}) as follows:
\begin{subequations}\bea
\ve^\a&=&\vf^\hf{\bm\ve}^\a~,
\\
\ve&=&\frac{1}{2\mub}\CD^\a\ve_\a
={1\over2\mub}\vf\bar{\vf}^{-1}\Big(
D^\a{\bm\ve}_\a + 2(D^\a\log{\vf}){\bm\ve}_\a\Big)~,
\non\\
&=&\frac{1}{2\mu}\CDB_\ad\bar{\ve}^\ad
={1\over2\mu}\bar{\vf}{\vf}^{-1}\Big(
\DB_\ad\bar{\bm\ve}^\ad + 2(\DB_\ad\log{\bar{\vf}})\bar{\bm\ve}^\ad\Big)
~.
\eea
\end{subequations}
The explicit expression for $\ve$ is
\bea
\ve&=&
-l
+{(2-\mu\qb^2)\e_\2\q
+(2-\mub\q^2)\bar{\e}^\2\qb
+\ri x^a(\mu\e_{\2}\s_a\qb
-\mub\q\s_a\bar{\e}^{\2})
+l(\mub\q^2+\mu\qb^2)\over\(1-{|\mu|^2\over 4}x^2\)}
\non\\
&&
+{
\mu\e_{\2}\q\qb^2
+\mub\q^2\bar{\e}^\2\qb 
+{\ri|\mu|^2\over 2}x^a(\e_{\2}\s_a\qb\q^2
-\q\s_a\bar{\e}^{\2}\qb^2)
+l|\mu|^2\q^2\qb^2
\over \(1-{|\mu|^2\over 4}x^2\)^2}
~.
\eea

As argued  earlier, the $\cN=2$ AdS transformation of the
prepotential $V_0$ is a pure gauge transformation.
Any AdS transformation should be accompanied by the inverse 
of the associated gauge transformation, in order to keep
$V_0$ fixed. This will result in modified supersymmetry transformations
of charged hypermultiplets (supersymmetry with central charge), 
in complete analogy with the rigid supersymmetric case \cite{vev}.
Here we provide  the expression for the induced gauge transformation of
$\hat{V}_0|=\re^{-(\sba-\s)J_{\1\2}}V_0|$, see eq. (\ref{V_0-reduced}).
A direct calculation gives
\bea
\d \hat{V}_0|=\l|+\tilde{\l}|~,
~~~~~~
\l|=\l_0|+\z\l_1|
+\z^2\l_2|
~,
\label{SUSY-var-V_0}
\eea
where
\bea
\l_0|&=&\ri{2{\e}_\2\q
-\ri\mub x^a_L\q\s_a\bar{\e}^\2
+l\mub\q^2
\over \(1-{|\mu|^2\over 4}x_L^2\)}
~,~~~
\l_1|=\L V_+~,~~~
\l_2|=\big(\ve^\a\CD_\a-\ve\mub\big)V_+~.
~~~~~~~~
\eea
Note that in eq.
(\ref{SUSY-var-V_0}), $\l_0|$ is chiral and $\l_1|$ can be seen to be complex linear, 
 $(\CDB^2-4\mu)\l_1=0$. This agrees with the requirement that the gauge parameter
 $\l$ should be  a weight-zero arctic  superfield.

\section{Dynamics in $\cN=2$ conformally flat superspace}
\setcounter{equation}{0}

In this section we study 
supersymmetric theories in an arbitrary 
conformally flat $\cN=2$ superspace $\cM^{4|8}$.
The corresponding covariant derivatives
$\cD_{\underline{A}}$ will be  assumed to have  the form 
(\ref{Finite-D}--\ref{Finite-D_c}), with $D_{\underline{A}}$
the covariant derivatives for ${\mathbb R}^{4|8}$.
It will also  be assumed that the torsion tensor $\cS_{ij}$ is real, 
$\cS_{ij} = {\bar \cS}_{ij}$.   The latter property means that 
$W_0:= {\rm e}^{-\s}$ is the field strength of an Abelian vector multiplet, 
that is the intrinsic vector multiplet for  $\cM^{4|8}$.

${}$For our subsequent consideration, 
it will be useful to view conformally flat $\cN=2$ supergeometry 
as a conformally flat $\cN=1$ superspace endowed with an Abelian $\cN=1$
vector multiplet. Indeed, for the 
covariant derivatives (\ref{Finite-D}--\ref{Finite-D_c}), it holds that 
\begin{subequations}
\bea
\re^{-(\sba-\s)|J_{\1\2}} \, \cD_\a^\1| \,\re^{(\sba-\s)|J_{\1\2}}
&=&\CD_\a + 2{\rm i}\, \cW_{0\a} J_{\2 \2}~,
\label{cd-N=1proj-al}       \\
\re^{-(\sba-\s)|J_{\1\2}} \,
\cDB^{\ad}_{ \1}| \, \re^{(\sba-\s)|J_{\1\2}}&=&\CDB^\ad
- 2{\rm i} \,{\bar \cW}_0^\ad J_{\1 \1}
\label{cd-N=1proj-ald} 
~.
\eea
\end{subequations}
Here the operators $\nabla_A = (\nabla_a, \CD_\a,\CDB^\ad)$ 
have the form (\ref{N=1der-al}--\ref{N=1der-a}), where 
$D_A = (D_a, D_\a,{\bar D}^\ad)$ are the flat $\cN=1$ covariant derivatives, 
and the chiral superfield $\vf$ is defined as 
\bea
\vf:=W_0|~,\qquad
\DB^\ad\vf=0~.
\label{vff}
\eea
The spinor superfield in (\ref{cd-N=1proj-al}), $\cW_0^\a$, 
is the covariantly chiral field strength of an Abelian $\cN=1$ vector multiplet, 
\be
{\bar \nabla}_\ad  \cW_0^\a= 0~, 
\qquad \nabla^\a  \cW_{0\a}= {\bar \nabla}_\ad  {\bar \cW}_0^\ad~.
\ee 
and  is related to $W_0$ as follows:
\be
\cW_{0\a} = \vf^{-3/2} W_{0\a}~, 
\qquad W_{0\a} := -\frac{\rm i}{2} D^{\2}_\a W_0|~.
\label{W-spinor-reduced}
\ee
In the case of $\cN=2$ AdS superspace, 
$\vf$ is given by eq. (\ref{vf}) and $\cW_{0\a} =0$.

In accordance with \cite{KLRT-M}, off-shell hypermultiplets 
are described by covariant arctic superfields of weight $n$, $\U^{(n)}(u^+)$, 
and their smile-conjugates. Given such a superfield in  $\cM^{4|8}$, 
we can use the standard  representation $\U^{(n)}(u^+)=(u^{+\1})^n \U^{[n]}(\z)$, 
and then 
\bea
\re^{-(\sba-\s) J_{\1\2}}\U^{[n]}(\z)\Big|=\re^{-\frac{n}{2}(\sba-\s)}\U^{[n]}
\big(\re^{(\sba-\s)}\z\big)\Big| 
\equiv \F+\z\G+\sum_{k=2}^{+\infty}\z^k \hat{\U}_k|~.
\eea
Here the leading components  $\F$ and $\G$ are covariantly chiral and complex linear, 
respectively,
\be
 \CDB^\ad\F=0~, \qquad \(\CDB^2-4R \)\G=0~,
\ee
where  $R=-(1/4)\vf^{-2}\DB^2\bar{\vf}$ 
is the chiral scalar component of the torsion
in the $\cN=1$ conformally flat superspace, 
see. e.g.   \cite{BK} for a review.

\subsection{Projecting the $\cN=2$ action  into $\cN=1$ superspace: II}

Our first goal is to project the supersymmetric action  
(\ref{InvarAc}) corresponding to   $\cM^{4|8}$,
\bea
S&=&
\frac{1}{2\pi} \oint_C (u^+ \rd u^{+})
\int \rd^4 x \, {\rm d}^4\q{\rm d}^4{\bar \q}\,\cE\, \frac{\cL^{++}}{(\cS^{++})^2}~,
\label{InvarAc1}
\eea
into $\cN=1$ superspace.
Using the super-Weyl transformation laws given in section 4,
for the superfields appearing in  (\ref{InvarAc1}) we find 
\bea
\cL^{++}&=& \re^{\s+\sba}L^{++}~,\qquad D_\a^+L^{++}=\DB^{+}_\ad L^{++}=0~,\non \\
\cS^{++}&=&\re^{\s+\sba}\S_0^{++}~, \qquad \cE=1~,
\eea
where 
\be
\S_0^{++} ={1\over 4}(D^+)^2W_0 = {1\over 4}(\DB^+)^2\bar{W}_0~.
\ee
The new Lagrangian, $L^{++}$, is a real weight-two projective multiplet 
in the flat $\cN=2$ superspace. 

In the action obtained, 
\bea
S&=&
\frac{1}{2\pi} \oint_C (u^+ \rd u^{+})
\int \rd^4 x \, {\rm d}^4\q{\rm d}^4{\bar \q}\, \frac{\re^{-\s-\sba}L^{++}}{(\S_0^{++})^2}~,
\label{InvarAc2}
\eea
we can make use of the identity
\bea
(D^+)^4\re^{-\s-\sba}=\Big({1\over 4}(D^+)^2W_0\Big)\Big({1\over 4}(\DB^+)^2\bar{W}_0\Big)
=(\S_0^{++})^2
~,~~~
(D^+)^4:={1\over 16}(D^+)^2(\DB^+)^2~,~~~~~~
\eea
and then transform (\ref{InvarAc2}) 
in the following way:
\bea
S&=&
\frac{1}{2\pi} \oint_C {(u^+ \rd u^{+})\over (u^+u^-)^4}
\int \rd^4 x \,(D^-)^4(D^+)^4\,\frac{\re^{-\s-\sba}L^{++}}{(\S_0^{++})^2}\Big|_{\q=0}
\non
\\
&=&
\frac{1}{2\pi} \oint_C {(u^+ \rd u^{+})\over (u^+u^-)^4}
\int \rd^4 x \,(D^-)^4\,L^{++}\Big|_{\q=0}~,
\label{InvarAc3}
\eea
where
\be
D^-_\a := u^-_i D^{i}_\a ~, \quad 
{\bar D}^-_\ad := u^-_i {\bar D}^{i}_\ad  ~, 
\qquad 
(D^-)^4:={1\over 16}(D^-)^2(\DB^-)^2~.
\ee
This action can be seen to be invariant 
under arbitrary projective transformations of the form
(\ref{projectiveGaugeVar}).
Without loss of generality, we can assume the north pole of ${\mathbb C}P^1$ 
to be  outside of the integration contour, hence $u^{+i}$
can be represented as $u^{+i}=u^{+\1}(1,\z)$,
with $\z$ the local complex coordinate for   ${\mathbb C}P^1$.
Using the projective invariance (\ref{projectiveGaugeVar}), we can then choose
$u^-_i$ to be  $u^-_i=(1,0)$.
${}$Finally, representing $L^{++}$ in the form
\be
L^{++}(z,u^+) =  {\rm i}\, u^{+\1} u^{+\2}\,
L(z,\z) =  {\rm i} \big( u^{+\1} \big)^2 \z\, L(z,\z)~, 
\ee
and also using the fact that $L^{++}$ enjoys the constraints
$\z_i D^i_\a L=\z_i {\bar D}^i_\ad L=0$, we can finally rewrite  
$S$ as an integral over the $\cN=1$ superspace
parametrized by the coordinates: $(x^a, \q^\a_{\1}, {\bar \q}^{\1}_\ad)$. 
The result is
\bea
S=
\frac{1}{2\pi \rm i}  \oint_C \frac{\rd \z}{\z}
\int \rd^4 x \, {\rm d}^2\q{\rm d}^2{\bar \q}\,  L(\z) \Big|~.
\label{InvarAc4}
\eea
As a last step, we replace here $L(\z)|$ with the $\cN=1$ projection 
of $\cL (\z)$ defined as 
$\cL^{++}(u^+)  =  {\rm i} \big( u^{+\1} \big)^2 \z\, \cL(\z)$. Thus
\bea
\cL (\z)|=\big( \re^{\s+\sba}L (\z)\big)\big| = \frac{1}{ \vf\vfb } L(\z)\big|~,
\eea 
and then  the  action obtained can be rewritten as
\bea
S=
\frac{1}{2\pi \rm i}  \oint_C \frac{\rd \z}{\z}
\int \rd^4 x \, {\rm d}^2\q{\rm d}^2{\bar \q}\, \vf\vfb \,\cL (\z)\big|
=
\frac{1}{2\pi \rm i}  \oint_C \frac{\rd \z}{\z}
\int \rd^4 x \, {\rm d}^2\q{\rm d}^2{\bar \q}\, E\, \cL (\z)\big|
~.
\label{InvarAc5}
\eea
This is the desired $\cN=1$ projection of the action (\ref{InvarAc1}).
In the AdS case, the above action  coincides with 
(\ref{InvarAc5-N=2-N=1}).

As follows from eqs. (\ref{cd-N=1proj-al}) and (\ref{cd-N=1proj-ald}), 
the projection into $\cN=1$ superspace should be accompanied 
by the U(1)-rotation $\re^{-(\sba-\s)|J_{\1\2}}$ applied to all superfields. 
This means that the final expression for the action (\ref{InvarAc5}) is 
\bea
S=
\frac{1}{2\pi \rm i}  \oint_C \frac{\rd \z}{\z}
\int \rd^4 x \, {\rm d}^2\q{\rm d}^2{\bar \q}\, E\, \cL \Big(\frac{\vf }{\bar \vf} \,\z\Big)\Big|
~.
\label{InvarAc6}
\eea
In the rest of this section, the  U(1)-rotation $\re^{-(\sba-\s)|J_{\1\2}}$
will be assumed to be performed.

\subsection{Massive hypermultiplets in AdS${}^{4|8}$}

As a simple application of the formalism developed, 
we consider the massive hypermultiplet model (\ref{massive-hyper}) 
in AdS${}^{4|8}$ 
(the massive model (\ref{massive-conformal- hyper}) can be studied similarly).
The corresponding Lagrangian to be used in (\ref{InvarAc6}) is
\bea
\cL|={1\over |\mu|}\bmS(\z){\widetilde{\bm \U}}(\z)|\re^{mV_0(\z)|}{\bm \U}(\z)|
\label{L-0}~,
\eea
We remind that  all the superfields are assumed to have been subjected to 
the U(1)-rotation $\re^{-(\sba-\s)|J_{\1\2}}$.

The weight-zero arctic superfield ${\bm\U}$ is characterized by the decomposition
(\ref{weight-zero}). For the prepotential $V_0$ of the intrinsic vector multiplet, we have
\bea
\re^{mV_0(\z)|}=\Big(1+m\z V_+\Big)\Big(1-{m\over\z}V_-\Big)
~,\quad
V_{+}=\ri\vf^2\bar{\vf}^{-1}\,\qb^2~,\quad V_{-}=-\ri\bar{\vf}^2{\vf}^{-1}\,\q^2~.~~~
\eea
It is then natural to generalize the superfield redefinition (\ref{U'}) to the massive case as
follows:
\bea
&&{\bm{\U}}'(\z)|:=\(1+\l\z\)\(1+\z V_+\){\bm\U}(\z)|
~,\qquad {\bm{\U}}' (\z) |={\bm\F}+\z{\bm\G}'+\sum_{k=2}^{\infty} {\bm\U}'_k \z^k~.
\eea
The component superfield ${\bm \G}'$ is now constrained by
\bea
-{1\over 4}\(\CDB^2-4\mu\){\bm \G}'&=&\ri |\m|  \(1+\frac{m}{|\m|}\){\bm\F}~,
\label{CNM-2}
\eea
while the components ${\bm\U}'_k$, $k>1$, are complex unconstrained.
Now, the contour integral in the action  generated by  the Lagrangian (\ref{L-0}) can easily  be performed, and the auxiliary fields  integrated out.
As a result, the action becomes 
\bea
S=\int \rd^4 x \, {\rm d}^2\q{\rm d}^2{\bar \q}\, E\Big(\bar{{\bm\F}}{\bm\F}-\bar{{\bm\G}}'{\bm\G}'\Big)
\label{S-CNM-massive}
~.
\eea
It is manifestly $\cN=1$ supersymmetric. 
It also possesses hidden second supersymmetry and U(1) symmetry. 
These are generated by a real parameter $\ve$ under the constraints (\ref{epsilon}),
and have the following form:
\bea
\d_\ve {\bm\F}&=&\ri\ve|\mu|\(1+\frac{m}{|\m|}\){\bm\F}-\(\bar{\ve}_\ad\CDB^\ad+\ve\mu\){\bm\G}'=
-{1\over 4}(\CDB^2-4\mu)(\ve{\bm\G}')
~, \non \\
\d_\ve {\bm\G}'&=&\ri\ve|\mu|\(1+\frac{m}{|\m|}\){\bm\G}'+\(\ve^\a\CD_\a+\ve\mub\){\bm\F}
~.
\eea
This transformation reduces to (\ref{SUSY-chiral-linear}) for $m=0$.
A purely chiral action, which is dual to (\ref{S-CNM-massive}),
proves to be
\bea
\int \rd^4 x \, {\rm d}^2\q{\rm d}^2{\bar \q}\, E\Big(
\bar{{\bm\F}}{\bm\F}
+\bar{{\bm\Psi}}{\bm\Psi}
-\ri\frac{\bar \mu}{|\mu |} \Big(1+\frac{m}{|\m|}\Big){\bm\Psi}{\bm\F} 
+\ri\frac{\mu}{|\mu |} \Big(1+\frac{m}{|\m|}\Big) {\bar {\bm\Psi}} {\bar {\bm\F}}
\Big)
~.
\label{S-CC-massive}
\eea
This action  reduces to (\ref{S-CNM}) for $m=0$.
Another interesting special case is $m =- |\m |$ for which (\ref{S-CC-massive}) turns into 
the superconformal massless action (\ref{conf-act-1-dual}).

The symmetry group of (\ref{S-CC-massive}) is OSp$(2|4)$.
The second SUSY and U(1) transformations are:
\bea
\d_\ve{\bm\F}=-{1\over 4}(\CDB^2-4\mu)(\ve\bar{\bm{\Psi}})~, \qquad
\d_\ve{\bm\Psi}={1\over 4}(\CDB^2-4\mu)(\ve\bar{\bm{\F}})~.
\label{SUSY-m}
\eea
Such transformations are $m$-independent and  identical to those 
which occur in the different models (\ref{conf-act-1-dual}).
This indicates that the transformations (\ref{SUSY-m}), in conjunction
with the $\cN=1$ AdS transformations, form  a closed algebra with a central charge proportional 
to $m$. 
This is indeed the case. One can check that transformations (\ref{SUSY-m}) have 
a manifestly $\cN=2$ supersymmetric realization. 
The latter is given in terms of an isospinor superfield
$q^i$ obeying the constraints 
\be
\cD^{(i}_\a q^{j)} = {\bar \cD}^{(i}_\ad q^{j)} = 0~
\ee
which generalize Sohnius' construction \cite{Sohnius}
for the off-shell hypermultiplet with intrinsic central charge \cite{Fayet}.
Unlike the arctic hypermultiplets (or more general harmonic $q^+$-hypermultiplets 
\cite{GIKOS,GIOS}), 
the above realization can only be used for the construction of simplest supersymmetric theories.

\subsection{Vector multiplet self-couplings}

We now turn our attention to the system of Abelian vector multiplets 
described by the Lagrangian
\bea
\cL^{++}=-\frac{1}{4} V_0 \Big{[} \Big(  (\cD^+)^2+4\cS^{++} \Big) \cF (\cW_I)
+\Big( (\cDB^+)^2+4\cS^{++}\Big){\bar \cF} ( \bar{\cW}_I) \Big{]}
~,~~~
\label{special-vector2}
\eea
In the AdS case, this Lagrangian becomes (\ref{special-vector}). 
Here we will consider the more general case of an arbitrary conformally flat superspace. 
 We are interested in reducing the model (\ref{special-vector2}) 
to $\cN=1$ conformally flat superspace.
Using conformal flatness, it turns out that the dynamics of (\ref{special-vector2}) 
is equivalently described by the Lagrangian
 \bea
L^{++}=-{1\over 4}V_0\Big{[}(D^+)^2
W_0\cF\({W_I\over W_0}\)
+(\DB^+)^2\bar{W}_0\cF\({\bar{W}_I\over \bar{W}_0}\)
\Big{]}
~,~~~
\label{special-vector-2}
\eea
where
\bea
&&\cW_I=W_0^{-1}W_I~, \qquad \DB_\ad^iW_I=0~, \qquad 
D^{ij}W_I=\DB^{ij}\bar{W}_I~.~~~
\label{useful-110}
\eea
For the general conformally flat supergeometry, 
the superfield $W_0=\re^{-\s}$ is only constrained to
obey the equation for the field strength of an Abelian vector multiplet 
in $\cN=2$ flat superspace, and otherwise it is arbitrary. 
The field strength $W_0$ is generated by 
a weight-zero tropical prepotential $V_0(\z)$, 
\bea
V_0(\z) =\sum_{k=-\infty}^{+\infty}\z^k v_k~, \qquad \overline{v_k}=(-1)^k v_{-k}~,
\qquad
D_\a^\1 v_k=D_\a^\2 v_{k+1}~,~~~~~~
\eea
The field strength is given as 
\bea 
W_0={\ri\over 4}\DB_\1^2 v_1={\ri\over 4}\DB_\2^2 v_{-1}~.
\label{useful-112}
\eea

The resulting flat-superspace action is
\bea
S=
\frac{\ri}{4} \oint_C \frac{\rd \z}{2\pi\ri\z}
\int\rd^4 x \, {\rm d}^2\q{\rm d}^2{\bar \q}\,
V_0(\z) \frac{\z_i\z_j}{\z}\Big{[}D^{ij}W_0\cF\Big({W_I\over W_0}\Big)
+\DB^{ij}\bar{W}_0\cF\Big({\bar{W}_I\over \bar{W}_0}\Big)
\Big{]}
\Big|~.~~~
\eea
It involves only 
the component superfieds $v_{-1},\, v_{0}$ and
$v_{1}$ of $V_0(\z)$.
Computing the contour integral, performing some 
$D$-algebra manipulations and using the identities
(\ref{useful-110}) and (\ref{useful-112}),
one can obtain the equivalent form for the action:
\bea
S&=&
\int \rd^4 x \, {\rm d}^4\q\, \vf\bar{\vf}\,
\bar{\F}_I\cF^I\(\F\)
+\int \rd^4 x \, {\rm d}^2\q\,
\vf^3\,R\big(2\cF\(\F\)-\F_I \cF^I\(\F\)\big)
\non\\
&&
+\int \rd^4 x \, {\rm d}^2\q\, \Big{[}
W^\a_0W_{0\a}\big(2\cF\(\F\)-2\F_I\cF^I\(\F\)+\F_I\F_J\cF^{IJ}(\F)\big)
\non\\
&&
+2W^\a_0W_{I\a}\big(\cF^I\(\F\)-\F_J\cF^{IJ}(\F)\big)
+W^\a_IW_{J\a }\cF^{IJ}(\F)
\Big{]}
~+~{\rm c.c.}~
\eea
Here we have introduced  the $\cN=1$ components, $\F_I$ and $W_{I\a}$, of $W_I$
defined as follows:
\bea
\vf \, \F_I=W_I|~,~~~W_{I\a}:=-{\ri\over 2} D_\a^\2W_I\big|~,~~~~~~
D^\a W_{I\a}=\DB_\ad\bar{W}_I^\ad~,
\eea
The similar components of 
$W_0^\a$ are defined  in  eqs. 
(\ref{vff}) and (\ref{W-spinor-reduced}).
Associated with $W_{I\a}$ is the curved-superspace field 
strength
$\cW_{\a I}=\vf^{-3/2}W_{\a I}$, 
which obeys the Bianchi identity
$\CDB^\ad\cW_{\a I}=0,\,\CD^\a\cW_{\a I}=\CDB_\ad\bar{\cW}^\ad_{ I}$.
In terms of the superfields introduced, the action takes the following final form:
\bea
S&=&
\int \rd^4 x \, {\rm d}^4\q\, E\,
\bar{\F}_I\cF^I\(\F\)
\non\\
&+&
\int \rd^4 x \, {\rm d}^4\q\, \frac{E}{R}\Big{[}
R\big(2\cF\(\F\)-\F_I \cF^I\(\F\)\big)
+\cW^\a_0\cW_{0\a}\big(2\cF\(\F\)-2\F_I\cF^I\(\F\)+\F_I\F_J\cF^{IJ}(\F)\big)
\non\\
&&
+2\cW^\a_0\cW_{I\a}\big(\cF^I\(\F\)-\F_J\cF^{IJ}(\F)\big)
+\cW^\a_I\cW_{J\a}\cF^{IJ}(\F)
\Big{]}
~+~{\rm c.c.}
\eea
If $\cF (\F)$ is a homogeneous function of degree two, $\F_I \cF^I\(\F\) =2\cF (\F)$,
the action considerably simplifies, in particular all dependence on $\cW^\a_0 $
disappears,
\bea
S&=&
\int \rd^4 x \, {\rm d}^4\q\, E\,
\bar{\F}_I\cF^I\(\F\)
+\int \rd^4 x \, {\rm d}^4\q\, \frac{E}{R}\cW^\a_I\cW_{J\a}\cF^{IJ}(\F)
~+~{\rm c.c.}
\eea
The action also simplifies drastically in the case of AdS${}^{4|8}$
where $\cW^\a_0 =0$.

\section{Open problems} 
\setcounter{equation}{0}

To conclude this paper, we would like to list a few interesting open problems.

It the  $\cN=1$ AdS supersymmetry, there exists a very nice classification 
of the off-shell superfield types due to Ivanov and Sorin \cite{IS} (see also
\cite{West} for a review), which is based on their local superprojectors. 
It would be interesting to carry out a similar analysis for the case of $\cN=2$ AdS superspace.
This might be useful for deriving  a manifestly $\cN=2$ supersymmetric formulation 
for the off-shell higher spin $\cN=2$ supermultiplets  \cite{GKS} on AdS${}^4$.

When realizing AdS${}^{4|8}$ as a conformally flat superspace, we 
used the stereographic coordinates for AdS${}^4$ (defined in Appendix D), 
in which the metric is manifestly SO(3,1) invariant. By analogy with the five-dimensional
consideration of \cite{KTM4}, it would be interesting  to re-do the whole analysis
in Poincar\'e parametrization\footnote{Similar to the stereographic coordinates,
these coordinates  cover 
one-half of the AdS hyperboloid.} 
in which the metric for AdS${}^4$ looks like 
\be
{\rm d}^2s = \Big(\frac{R}{z}\Big)^2 \Big( 
\eta_{\hat{m}\hat{n}}\,{\rm d}x^{\hat{m}}   {\rm d}x^{\hat{n}} +  {\rm d}z^2 \Big)~, \qquad 
R={\rm const}~, \qquad 
\hat{m},\hat{n} =0, 1,2,~
\ee
with $\eta_{\hat{m}\hat{n}}$ the three-dimensional Minkowski metric.
First of all, this would give direct access to three-dimensional superconformal 
theories. Second, the Poincar\'e coordinates should be very useful
for the explicit elimination of the auxiliary superfields in nonlinear sigma-models of the form 
(\ref{non-conf-sigma}), see \cite{KTM4} for more detail.

It would be desirable to develop harmonic-superspace techniques for AdS${}^{4|8}$.
This should proceed similarly to the harmonic-superspace construction
developed  in the case of 5D $\cN=1$ AdS superspace \cite{KT-M}.
The harmonic superspace approach is known to be most suitable for quantum calculations 
in $\cN=2$ super Yang-Mills theories. Thus it would be very interesting, {\it e.g.}, to see 
how the covariant harmonic supergraphs \cite{BBKO,BKO} generalize to the AdS case.
\\

\noindent
{\bf Acknowledgements:}\\
We are grateful to Darren Grasso for reading the manuscript.
We thank the organizers of the 2008 Simons Workshop in Mathematics and Physics,
where this project was completed, for their hospitality.
This work is supported  in part by the Australian Research Council.

\appendix

\section{Superspace geometry of conformal supergravity} 
\setcounter{equation}{0}
\label{SCG}

Consider a curved 4D $\cN=2$ superspace  $\cM^{4|8}$ parametrized by 
local bosonic ($x$) and fermionic ($\q, \bar \q$) 
coordinates  $z^{\underline{M}}=(x^{m},\q^{\mu}_i,{\bar \q}_{\dot{\mu}}^i)$,
where $m=0,1,\cdots,3$, $\mu=1,2$, $\dot{\mu}=1,2$ and  $i=\1,\2$.
The Grassmann variables $\q^{\mu}_i $ and $\teb_{\dot{\mu}}^i$
are related to each other by complex conjugation: 
$\overline{\q^{\mu}_i}=\teb^{\dot{\mu}i}$. 
The structure group is chosen to be ${\rm SO}(3,1)\times {\rm SU}(2)$ \cite{Grimm,KLRT-M},
and the covariant derivatives 
$\cD_{\underline{A}} =(\cD_{{a}}, \cD_{{\a}}^i,\cDB^\ad_i)$
have the form 
\bea
\cD_{\underline{A}}&=&\cE_{\underline{A}}
~+~\O_{\underline{A}}
~+~\F_{\underline{A}}~.
\label{CovDev}
\eea
Here $\cE_{\underline{A}}=\cE_{\underline{A}}{}^{\underline{M}}(z)\pa_{\underline{M}}$ 
is the supervielbein, with $\pa_{\underline{M}}=\pa/\pa z^{\underline{M}}$,
\bea
\O_{\underline{A}}&=&\hf\O_{\underline{A}}{}^{bc}M_{bc}=\O_{\underline{A}}{}^{\b\g}\,M_{\b\g}
+{\bar \O}_{\underline{A}}{}^{\bd\gd}\,\bar{M}_{\bd\gd}
\label{Lorentzconnection}
\eea
is the Lorentz connection,
\bea
\F_{\underline{A}}=\F_{\underline{A}}{}^{kl}J_{kl}~,~~~J_{kl}=J_{lk}
\label{SU(2)connection}
\eea
is the SU(2)-connection.
The Lorentz generators with vector indices ($M_{ab}=-M_{ba}$) and spinor indices
($M_{\a\b}=M_{\b\a}$ and ${\bar M}_{\ad\bd}={\bar M}_{\bd\ad}$) are related to each other 
by the rule:
$$
M_{ab}=(\s_{ab})^{\a\b}M_{\a\b}-(\tilde{\s}_{ab})^{\ad\bd}\bar{M}_{\ad\bd}~,~~~
M_{\a\b}=\hf(\s^{ab})_{\a\b}M_{ab}~,~~~
\bar{M}_{\ad\bd}=-\hf(\tilde{\s}^{ab})_{\ad\bd}M_{ab}~.
$$ 
The generators of SO(3,1)$\times$SU(2)
act on the covariant derivatives as follows:\footnote{In what follows, 
the (anti)symmetrization of $n$ indices 
is defined to include a factor of $(n!)^{-1}$.}
\bea
&{[}J_{kl},\cD_{\a}^i{]}
=-\d^i_{(k}\cD_{\a l)}~,
\qquad
{[}J_{kl},\cDB^{\ad}_i{]}
=-\ve_{i(k}\cDB^\ad_{l)}~, \non \\
&{[}M_{\a\b},\cD_{\g}^i{]}
=\ve_{\g(\a}\cD^i_{\b)}~,\qquad
{[}\bar{M}_{\ad\bd},\cDB_{\gd}^i{]}=\ve_{\gd(\ad}\cDB^i_{\bd)}~,
~~~
{[}M_{ab},\cD_c{]}=2\eta_{c[a}\cD_{b]}~,
\label{generators}
\eea
while 
${[}M_{\a\b},\cDB_{\gd}^i{]}=
{[}\bar{M}_{\ad\bd},\cD_{\g}^i{]}={[}J_{kl},\cD_a{]}=0$.
Our notation and conventions correspond to \cite{BK,KLRT-M}; they 
almost coincide with 
those used in \cite{WB} except for the normalization of the 
Lorentz generators, including a sign in the definition of  
the sigma-matrices $\s_{ab}$ and $\tilde{\s}_{ab}$.

The supergravity gauge group is generated by local transformations
of the form 
\be
\d_K \cD_{\underline{A}} = [K  , \cD_{\underline{A}}]~,
\qquad K = K^{\underline{C}}(z) \cD_{\underline{C}} +\hf K^{ c  d}(z) M_{c  d}
+K^{kl}(z) J_{kl}  ~,
\label{tau}
\ee
with the gauge parameters
obeying natural reality conditions, but otherwise  arbitrary. 
Given a tensor superfield $U(z)$, with its indices suppressed, 
it transforms as follows:
\bea
\d_K U = K\, U~.
\label{tensor-K}
\eea

The  covariant derivatives obey (anti-)commutation relations of the form:
\bea
{[}\cD_{\underline{A}},\cD_{\underline{B}}\}&=&
\cT_{ \underline{A}\underline{B} }{}^{\underline{C}}\cD_{\underline{C}}
+\hf \cR_{\underline{A} \underline{B}}{}^{{c}{d}}M_{{c}{d}}
+\cR_{ \underline{A} \underline{B}}{}^{kl}J_{kl}
~,
\label{algebra}
\eea
where $\cT_{\underline{A} \underline{B} }{}^{\underline{C}}$ is the torsion, 
and $\cR_{\underline{A} \underline{B}}{}^{kl}$ and 
$\cR_{ \underline{A} \underline{B}}{}^{{c}{d}}$ 
constitute the curvature.
${}$The torsion is subject to the 
following constraints \cite{Grimm}:
\bea
&\cT_{\a}^i{}_{\b}^j{}^{c}=
\cT_{\a}^i{}_{\b}^j{}^{\g}_k=\cT_{\a}^i{}_{\b}^j{}_{\dot{\g}}^k
=\cT_{\a}^i{}^{\dot{\b}}_j{}^{\g}_k
=\cT_{a}{}_{\b}^j{}^c
=\cT_{ab}{}^{c}=0~,
\non\\
&\cT_{\a}^i{}^\bd_j{}^c=-2\ri\d^i_j(\s^{c})_{\a}{}^{\bd}~,~~~
\cT_{a}{}_{\b}^j{}^\g_k=\d^j_k\,\cT_{a\b}{}^\g~.
\label{constr-1}
\eea
Here we have omitted some constraints which follow by complex conjugation.
The algebra of covariant derivatives is \cite{KLRT-M}
\begin{subequations} 
\bea
\{\cD_\a^i,\cD_\b^j\}&=&
4\cS^{ij}M_{\a\b}
+2\ve^{ij}\ve_{\a\b}\cY^{\g\d}M_{\g\d}
+2\ve^{ij}\ve_{\a\b}\bar{\cW}^{\gd\dd}\bar{M}_{\gd\dd}
\non\\
&&
+2 \ve_{\a\b}\ve^{ij}\cS^{kl}J_{kl}
+4 \cY_{\a\b}J^{ij}~,
\label{acr1} \\
\{\cD_\a^i,\cDB^\bd_j\}&=&
-2\ri\d^i_j(\s^c)_\a{}^\bd\cD_c
+4\d^{i}_{j}\cG^{\d\bd}M_{\a\d}
+4\d^{i}_{j}\cG_{\a\gd}\bar{M}^{\gd\bd}
+8 \cG_\a{}^\bd J^{i}{}_{j}~,
\label{acr2}
\\
{[}\cD_a,\cD_\b^j{]}&=&
\ri(\s_a)_{(\b}{}^{\bd}\cG_{\g)\bd}\cD^{\g j}
+{\ri\over 2}\Big(({\s}_a)_{\b\gd}\cS^{jk}
-\ve^{jk}({\s}_a)_\b{}^{\dd}\bar{\cW}_{\dd\gd}
-\ve^{jk}({\s}_a)^{\a}{}_\gd \cY_{\a\b}\Big)\cDB^\gd_k
\non\\
&&
\qquad \qquad 
~+~\mbox{curvature terms}~.
\label{acr3}
\eea
\end{subequations}
Here the real four-vector $\cG_{\a \ad} $,
the complex symmetric  tensors $\cS^{ij}=\cS^{ji}$, $\cW_{\a\b}=\cW_{\b\a}$, 
$\cY_{\a\b}=\cY_{\b\a}$ and their complex conjugates 
$\bar{\cS}_{ij}:=\overline{\cS^{ij}}$, $\bar{\cW}_{\ad\bd}:=\overline{\cW_{\a\b}}$,
$\bar{\cY}_{\ad\bd}:=\overline{\cY_{\a\b}}$ obey additional differential constraints implied 
by the Bianchi identities \cite{Grimm,KLRT-M}.
Of special importance are
the following dimension 3/2 identities: 
\bea
\cD_{\a}^{(i}\cS^{jk)}= {\bar \cD}_{\ad}^{(i}\cS^{jk)}=0~.
\label{S-analit}
\eea

\section{Vector multiplets in conformal supergravity}
\setcounter{equation}{0}

Here we discuss the projective-superspace description 
of off-shell vector multiplets in 4D $\cN=2$ conformal supergravity.
Following the conventions adopted in   \cite{KLRT-M}, an Abelian vector multiplet
is described by  its field strength $\cW(z)$ which is covariantly chiral \be
{\bar \cD}^{\ad }_i \cW =0~,
\ee
and obeys  the Bianchi identity
\bea
\S^{ij}:=\frac{1}{4}\Big(\cD^{\g(i}\cD_\g^{j)}+4\cS^{ij}\Big)\cW
&=&
\frac{1}{4}\Big(\cDB_\gd^{(i}\cDB^{ j) \gd}+ 4\bar{\cS}^{ij}\Big)\bar{\cW}=:{\bar \S}^{ij}
~.
\label{vectromul}
\eea
Under the infinitesimal super-Weyl transformations,   $\cW$ varies as 
\be
\d_{\s} \cW = \s \cW~.
\label{Wsuper-Weyl}
\ee
The super-Weyl transformation of $\S^{ij}$ is 
\be
\d_{\s} \S^{ij} = \big(\s +\bar \s \big)  \S^{ij} ~.
\label{S(++)super-Weyl}
\ee

The vector multiplet can also be described by its gauge field $\cV(z,u^+)$ which is 
a covariant real weight-zero tropical supermultiplet possessing 
the following expansion in the north chart of ${\mathbb C}P^1$:  
\bea
\cD^+_{\a} \cV  = {\bar \cD}^+_{\ad} \cV  =0~, \qquad
\cV(z,u^+)= 
\cV(z,\z) =\sum_{k=0}^{+\infty}\z^k \,\cV_k(z)~,\quad \cV_k=(-1)^k\bar{\cV}_{-k}~.~~
\eea
It turns out that the field strength $\cW$ and its conjugate $\bar \cW$ are expressed
in terms of the prepotential $V$ as follows:
 \begin{subequations} 
\bea
\cW (z)&=&
-{1\over 8\pi}\oint{(u^+\rd u^+)\over(u^+u^-)^2}(\cDB_{\ad}^-\cDB^{\ad-}+4\bar{\cS}^{--})\cV(z,u^+)
~,~~~  
\label{W}\\
\bar{\cW}(z) &=&-{1\over 8\pi}\oint{(u^+\rd u^+)\over(u^+u^-)^2}
(\cD^{\a-}\cD_{\a}^{-}+4\cS^{--})\cV(z,u^+)
~,~~~~~~~~~
\label{W-bar}
\eea
\end{subequations} 
with the contour integral being carried out around the origin.
These expressions can be shown to be invariant 
under arbitrary projective transformations of the form:
\be
(u_i{}^-\,,\,u_i{}^+)~\to~(u_i{}^-\,,\, u_i{}^+ )\,R~,~~~~~~R\,=\,
\left(\begin{array}{cc}a~&0\\ b~&c~\end{array}\right)\,\in\,{\rm GL(2,\mathbb{C})}~.
\label{projectiveGaugeVar}
\ee
Using the fact that $\cV(z,u^+)$ is a covariant projective supermultiplet,
$\cD^+_{\a} \cV  = {\bar \cD}^+_{\ad} \cV  =0$, one can show that the right-hand side 
of (\ref{W}) is covariantly chiral. For this, it is advantageous to make use of
the following equivalent representations:
\bea
\cW&=&{1\over 8\pi}\oint{\rd \z\over\z^2}\(\cDB_{\ad\1}\cDB^{\ad}_\1
+4\bar{\cS}_{\1\1}\)\cV(\z)
={\ri\over 4}\(\cDB_{\ad\1}\cDB^{\ad}_\1+4\bar{\cS}_{\1\1}\)\cV_1
~,~~~ 
\non \\
\cW
&=&{1\over 8\pi}\oint{\rd \z}(\cDB_{\ad\2}\cDB^{\ad}_\2+4\bar{\cS}_{\2\2})\cV(\z)
={\ri\over 4}\(\cDB_{\ad\2}\cDB^{\ad}_{\2} +4\bar{\cS}_{\2\2}\)\cV_{-1}~.
\label{W-V-1}
\eea

The field strength (\ref{W}) can be shown to be invariant under gauge transformations
of the form 
\be
\d \cV =\l  + \widetilde{\l}~,
\ee
with the gauge parameter $\l(z,u^+)$ being a covariant weight-zero arctic multiplet, 
and $\widetilde{\l}$ its smile-conjugate, 
 \begin{subequations} 
\bea
\cD^+_{\a} \l  &=& {\bar \cD}^+_{\ad} \l  =0~, \qquad
\l(z,u^+) = \l(z,\z) =\sum_{k=0}^{+\infty}\z^k\l_k(z) ~,  \\
\cD^+_{\a} \widetilde{\l}  &= &{\bar \cD}^+_{\ad} \widetilde{\l}  =0~, \qquad
\widetilde{\l}(z,u^+) = \widetilde{\l}(z,\z) =\sum_{k=0}^{+\infty}(-1)^k \z^{-k}\bar{\l}_k (\z)
~.
\eea
 \end{subequations} 
To prove the gauge invariance of $\cW$, the only non-trivial observation required is that 
the constraints on $\l$ and $\widetilde{ \l }$ imply 
\bea
\(\cDB_{\ad\1}\cDB^{\ad}_\1+4\bar{\cS}_{\1\1}\)\l_1=0~, \qquad 
\(\cDB_{\ad\2}\cDB^{\ad}_\2+4\bar{\cS}_{\2\2}\){\bar \l}_{1}=0~.
\eea

It can also be demonstrated that the following super-Weyl transformation of the gauge 
prepotential $\cV(z,u^+)$, 
\be
\d_\s \cV =0~,
\ee
implies the super-Weyl transformation of $\cW$, 
eq. (\ref{Wsuper-Weyl}).

\section{$\cN=1$ AdS Killing supervectors}
\setcounter{equation}{0}

The covariant derivatives of  the $\cN=1$ anti-de Sitter superspace AdS$^{4|4}$, 
\bea
\CD_A=(\CD_a, \CD_\a,\CDB^\ad)=E_A{}^{M}\pa_M+\hf\f_A{}^{bc}M_{bc}~,
\eea
obey the following (anti-)commutation relations:
\begin{subequations}
\bea
&&\{\CD_\a,\CD_\b\}=
-4\bar{\mu}M_{\a\b}~, \qquad ~~~
\{\CD_\a,\CDB^\bd\}=-2\ri(\s^c)_\a{}^\bd\CD_c
~,~~
\label{N=1-AdS-algebra-1}
\\
&&
{[}\CD_a,\CD_\b{]}=
-\frac{\ri}{ 2}   \mub ({\s}_a)_{\b\gd}\CDB^\gd~,\qquad
{[}\CD_a,\CD_b{]}=-|\mu|^2M_{ab}~,~~~~~~
\label{N=1-AdS-algebra-2}
\eea
\end{subequations}
with $\m$ a complex non-vanishing parameter which can be viewed to be  
a square root of the curvature of the anti-de Sitter space, 
see, e.g., \cite{BK} for more detail.
The symmetries of AdS$^{4|4}$ are generated by the corresponding Killing supervectors 
defined as 
\bea
\L=\l^a\CD_a+\l^\a\CD_\a+\lb_\ad\CDB^\ad~, \qquad
{[}\L+\o^{bc}M_{bc},\CD_A{]}=0~,
\label{N=1-killings-0}
\eea
for some local Lorentz transformation associated with  $\o^{bc}$. 
As shown in \cite{BK}, 
the equations in (\ref{N=1-killings-0}) are equivalent to 
\bea
&&\o_{\a\b}=\CD_\a\l_\b~,\qquad
\CD_\a\l^\a=0~,
\qquad
0={\ri\over 2}{\mu}\l_{\a\ad}
+\CDB_\ad\l_\a~,
\label{4-SK-1}
\\
&&0=\CD_{(\a}\l_{\b)\bd}~, \qquad
0=\CDB^\bd\l_{\a\bd}
+8\ri\l_\a~.
\label{4-SK-2}
\eea

\section{Stereographic projection for AdS spaces}
\setcounter{equation}{0}

Consider a $d$-dimensional anti-de Sitter space AdS${}_d$. 
It can be realized as a hypersurface in ${\mathbb R}^{d-1,2}$ parametrized
by Cartesian coordinates $Z^{\hat a} = (Z^d, Z^a)$, with $a=0,1, \dots, d-1$. 
The hypersurface looks like 
\be
-(Z^d)^2 -(Z^0)^2 +\sum_{i=1}^{d-1} (Z^i)^2 =  -(Z^d)^2 +Z^a Z_a = -R^2 ={\rm const}~.
\ee

One can introduce unconstrained local coordinates for AdS${}_d$ 
as a natural  generalization of the stereographic projection for $S^d$. 
Let us cover AdS${}_d$ by two charts: 
(i) the north chart in which 
$Z^d> -R$; and 
(ii) the south chart in which $Z^d<R$.
Given a point $Z^{\hat a}$ in the north chart, its local coordinates $x^a$ will be 
chosen to correspond to the intersection of the plane $Z^d=0$ 
and the straight line connecting $Z^{\hat a}$ and the ``north pole'' 
$Z^{\hat a}_{\rm north} = (-R, 0, \dots , 0)$. 
Similarly, given a point $Z^{\hat a}$ in the south chart, its local coordinates $y^a$ will be 
chosen to correspond to the intersection of the plane $Z^d=0$ 
with  the straight line connecting $Z^{\hat a}$ and the ``south pole'' 
$Z^{\hat a}_{\rm south} = (R, 0, \dots , 0)$. 

In the north chart, one finds 
\be
x^a = \frac{R \,Z^a}{R+Z^d}~, 
\qquad x^a x_a <R^2~. 
\ee
A short calculation for the induced  metric, ${\rm d}s^2 = -({\rm d} Z^d)^2 +{\rm d}Z^a \,{\rm d}Z_a $,
gives the conformally flat form:
\be
{\rm d}s^2 = \frac{4\, {\rm d}x^a {\rm d}x_a ~}{\big(1- R^{-2} \,x^2\big)^2} 
~, \qquad x^2 = x^bx_b~.
\label{metric-north}
\ee

In the south chart, one similarly gets 
\be
y^a = \frac{R \,Z^a}{R-Z^d}~, 
\qquad y^a y_a <R^2~. 
\ee
The metric is obtained from  (\ref {metric-north})  by replacing 
$x^a \to y^a$.

In the intersection of the two charts, the transition functions are:
\be
y^a = -R^2 \,\frac{x^a}{x^2}~.
\ee
This is an inversion, that is, a discrete conformal  transformation.


\begin{thebibliography}{10}
\small
\setlength{\parskip}{0pt}
\setlength{\itemsep}{0pt}

\bibitem{KLRT-M}
S.~M.~Kuzenko, U.~Lindstr\"om, M.~Ro\v cek and G.~Tartaglino-Mazzucchelli,
``4D N=2 supergravity and projective superspace,'' arXiv:0805.4683.

\bibitem{KT-Msugra1}
S.~M.~Kuzenko and G.~Tartaglino-Mazzucchelli,
  ``Five-dimensional superfield supergravity,''
    Phys.\ Lett.\ B {\bf 661}, 42 (2008)
 [arXiv:0710.3440];
  ``5D supergravity and projective superspace,''
  JHEP {\bf 0802}, 004 (2008) [arXiv:0712.3102].

\bibitem{KT-Msugra3}
  S.~M.~Kuzenko and G.~Tartaglino-Mazzucchelli,
  ``Super-Weyl invariance in 5D supergravity,''
  JHEP {\bf 0804}, 032 (2008)
  [arXiv:0802.3953].

\bibitem{Zumino78} B. Zumino, ``Supergravity and Superspace,'' in 
{\it Recent Developments in Gravitation}, Carg\`ese 1978, 
M. L\'evy and S. Deser (Eds.), Plenum Press, New York, 1979, p. 405.

\bibitem{SG}
  W.~Siegel,
  ``Solution to constraints in Wess-Zumino supergravity formalism,''
  Nucl.\ Phys.\  B {\bf 142}, 301 (1978);
W.~Siegel and S.~J.~Gates, Jr.
  ``Superfield supergravity,''
  Nucl.\ Phys.\  B {\bf 147}, 77 (1979).

\bibitem{GGRS}
S.~J.~Gates, Jr., M.~T.~Grisaru, M.~Ro\v{c}ek and W.~Siegel,
{\it Superspace, Or One Thousand 
and One Lessons in Supersymmetry},
Benjamin/Cummings (Reading, MA),  1983 [hep-th/0108200].


\bibitem{BK}
I.~L. Buchbinder and S.~M. Kuzenko, {\it Ideas and Methods of Supersymmetry and
Supergravity, Or a Walk Through Superspace}, IOP, Bristol, 1998.

\bibitem{WZ-s}
J.~Wess and B.~Zumino,
``Superspace formulation of supergravity,''
Phys.\ Lett.\  B {\bf 66}, 361 (1977);
R.~Grimm, J.~Wess and B.~Zumino,
``Consistency checks on the superspace 
formulation of supergravity,''
Phys.\ Lett.\ B {\bf 73}, 415 (1978);
J.~Wess and B.~Zumino,
``Superfield Lagrangian for supergravity,''
Phys.\ Lett.\ {\bf B74},  51 (1978).

\bibitem{old}
K.~S.~Stelle and P.~C.~West,
``Minimal auxiliary fields for supergravity,''
Phys.\ Lett.\ {\bf B74}, 330 (1978);
S.~Ferrara and P.~van Nieuwenhuizen,
``The auxiliary fields of supergravity,''
Phys.\ Lett.\ {\bf B74}, 333 (1978).

 \bibitem{WB} J.~Wess and J.~Bagger,
{\it Supersymmetry and Supergravity},
Princeton Univ. Press, 1992.

\bibitem{KLR}
A. Karlhede, U. Lindstr\"om and M. Ro\v cek,
``Self-interacting tensor multiplets in N=2 superspace,''
Phys.\ Lett.\ B {\bf 147}, 297 (1984).

\bibitem{LR}
  U.~Lindstr\"om and M.~Ro\v{c}ek,
 ``New hyperk\"ahler  metrics  and new supermultiplets,''
  Commun.\ Math.\ Phys.\  {\bf 115}, 21 (1988);
  ``N=2 super Yang-Mills theory in projective superspace,''
  Commun.\ Math.\ Phys.\   {\bf 128}, 191 (1990).
 
 \bibitem{GKLR}
  S.~J.~Gates, Jr., A.~Karlhede, U.~Lindstr\"om and M.~Ro\v{c}ek,
  ``N=1 superspace components of extended supergravity,''
  Class.\ Quant.\ Grav.\  {\bf 1}, 227 (1984);
  ``N=1 superspace geometry of extended supergravity,''
  Nucl.\ Phys.\  B {\bf 243}, 221 (1984);
 J.~M.~F.~Labastida, M.~Ro\v{c}ek, E.~Sanchez-Velasco and P.~Wills,
  ``N=2 supergravity action in terms of N=1 superfields,''
  Phys.\ Lett.\  B {\bf 151}, 111 (1985);
 J.~M.~F.~Labastida, E.~Sanchez-Velasco and P.~Wills,
  ``The N=2 vector multiplet coupled to supergravity in N=1 superspace,''
  Nucl.\ Phys.\  B {\bf 256}, 394 (1985).
 
 \bibitem{Labastida}
  J.~M.~F.~Labastida, E.~Sanchez-Velasco and P.~Wills,
  ``N=2 conformal supergravity in N=1 superspace,''
  Nucl.\ Phys.\  B {\bf 278}, 851 (1986).
 
 \bibitem{FV}
  E.~S.~Fradkin and M.~A.~Vasiliev,
  ``Minimal set of auxiliary fields and S-matrix for extended supergravity,''
  Lett.\ Nuovo Cim.\  {\bf 25} (1979) 79;
 ``Minimal set of auxiliary fields in SO(2) extended supergravity,''
  Phys.\ Lett.\  B {\bf 85} (1979) 47;
 B.~de Wit and J.~W.~van Holten,
  ``Multiplets of linearized SO(2) supergravity,''
  Nucl.\ Phys.\  B {\bf 155}, 530 (1979);
 B.~de Wit, J.~W.~van Holten and A.~Van Proeyen,
  ``Transformation rules of N=2 supergravity multiplets,''
  Nucl.\ Phys.\  B {\bf 167}, 186 (1980).
 
\bibitem{BS}
  P.~Breitenlohner and M.~F.~Sohnius,
  ``Superfields, auxiliary fields, and tensor calculus for N=2 extended supergravity,''
  Nucl.\ Phys.\  B {\bf 165}, 483 (1980);
L.~Castellani, P.~van Nieuwenhuizen and S.~J.~Gates, Jr.,
  ``The constraints for N=2 superspace from extended supergravity 
in ordinary  space,''
  Phys.\ Rev.\  D {\bf 22}, 2364 (1980); 
 S.~J.~Gates, Jr.,
  ``Another solution for N=2 superspace Bianchi identities,''
  Phys.\ Lett.\  B {\bf 96}, 305 (1980);
 S.~J.~Gates, Jr. and W.~Siegel,
  ``Linearized N=2 superfield supergravity,''
  Nucl.\ Phys.\  B {\bf 195}, 39 (1982).
 
 \bibitem{Howe}
  P.~S.~Howe,
  ``Supergravity in superspace,''
  Nucl.\ Phys.\  B {\bf 199}, 309 (1982).
 
\bibitem{IS}
  E.~A.~Ivanov and A.~S.~Sorin,
  ``Superfield formulation of OSp(1,4) supersymmetry,''
  J.\ Phys.\ A  {\bf 13}, 1159 (1980).

\bibitem{Keck}
  B.~W.~Keck,
 ``An alternative class of supersymmetries,''
J.\ Phys.\ A  {\bf 8}, 1819 (1975).

\bibitem{Zumino}
  B.~Zumino, ``Nonlinear realization of supersymmetry in de Sitter space,''
 Nucl.\ Phys.\  B {\bf 127}, 189 (1977).

\bibitem{GKS}
S.~J.~Gates, Jr., S.~M.~Kuzenko and A.~G.~Sibiryakov,
``N = 2 supersymmetry of higher superspin massless theories,''
Phys.\ Lett.\  B {\bf 412}, 59 (1997) [hep-th/9609141];
  ``Towards a unified theory of massless superfields of all superspins,''
  Phys.\ Lett.\  B {\bf 394}, 343 (1997)
  [hep-th/9611193].
  
 \bibitem{KS}
  S.~M.~Kuzenko and A.~G.~Sibiryakov,
  ``Free massless higher superspin superfields on the anti-de Sitter
  superspace,''
  Phys.\ Atom.\ Nucl.\  {\bf 57} (1994) 1257.

\bibitem{SS}
A.~Y.~Segal and A.~G.~Sibiryakov,
``Explicit N = 2 supersymmetry for higher-spin massless fields in D = 4  AdS
superspace,''  Int.\ J.\ Mod.\ Phys.\  A {\bf 17}, 1207 (2002) [hep-th/9903122].

 \bibitem{GIKOS}
A.~S.~Galperin, E.~A.~Ivanov, S.~N.~Kalitsyn, V.~Ogievetsky, E.~Sokatchev, 
``Unconstrained N=2 matter, Yang-Mills and supergravity theories in harmonic
superspace,''
Class.\ Quant.\ Grav.\  {\bf 1}, 469 (1984).
 
\bibitem{GIOS}
A.~S.~Galperin, E.~A.~Ivanov, V.~I.~Ogievetsky and E.~S.~Sokatchev,
{\it Harmonic Superspace}, Cambridge University Press,  2001.
  
\bi{KT-M}
S.~M. Kuzenko and G. Tartaglino-Mazzucchelli,
``Five-dimensional N=1 AdS superspace:
Geometry,  off-shell multiplets and dynamics,''
Nucl. Phys. B {\bf 785}, 34 (2007), [arXiv:0704.1185].

\bi{Rosly}
 A.~A.~Rosly,
``Super Yang-Mills  constraints 
as integrability conditions,'' in {\it Proceedings of the International 
Seminar on Group Theoretical 
Methods in Physics},'' (Zvenigorod, USSR, 1982),
M. A. Markov  (Ed.), 
Nauka, Moscow, 1983, Vol. 1, p. 263 (in Russian);
{\it see also}  A.~A.~Rosly and A.~S.~Schwarz,
``Supersymmetry in a space with auxiliary dimensions,''
Commun.\ Math.\ Phys.\  {\bf 105}, 645 (1986).

 \bibitem{K-hyper1}
 S.~M.~Kuzenko,
 ``On compactified harmonic/projective superspace, 5D superconformal
 theories, and all that,''
  Nucl.\ Phys.\  B {\bf 745}, 176 (2006)
  [hep-th/0601177]. 
  
  \bibitem{K-hyper2}
S.~M.~Kuzenko, ``On superconformal projective hypermultiplets,''
JHEP {\bf 0712}, 010 (2007) [arXiv:0710.1479].

\bibitem{G-RLRvUW}
F.~Gonzalez-Rey, U.~Lindstr\"om
M.~Ro\v{c}ek, R.~von Unge and S.~Wiles, 
 ``Feynman rules in N = 2 projective superspace. 
I: Massless  hypermultiplets,''
  Nucl.\ Phys.\ B {\bf 516}, 426 (1998) {[hep-th/9710250]}.

\bibitem{GS}
S.~J.~Gates, Jr. and W.~Siegel,
  ``Variant superfield representations,''
  Nucl.\ Phys.\  B {\bf 187}, 389 (1981).

\bibitem{LR2}
  U.~Lindstr\"om and M.~Ro\v{c}ek,
  ``Scalar tensor duality and $N=1, 2$ nonlinear sigma models,''
  Nucl.\ Phys.\  B {\bf 222}, 285 (1983).

\bibitem{DG}
 B.~B.~Deo and S.~J.~Gates, Jr.,
  ``Comments on nonminimal N=1 scalar multiplets,''
  Nucl.\ Phys.\  B {\bf 254}, 187 (1985).

\bibitem{deWPV}
B.~de Wit, R.~Philippe and A.~Van Proeyen,
``The improved tensor multiplet in N = 2 supergravity,''
Nucl.\ Phys.\ B {\bf 219}, 143 (1983).

\bibitem{GIO}
A.~Galperin, E.~Ivanov and V.~Ogievetsky,
``Superspace actions and duality transformations 
for N = 2 tensor multiplets,''
Sov.\ J.\ Nucl.\ Phys.\  {\bf 45}, 157 (1987);
``Duality transformations and most general matter 
self-coupling in N = 2 supersymmetry,''
Nucl.\ Phys.\ B {\bf 282}, 74 (1987).

\bibitem{deWR}
  B.~de Wit and M.~Ro\v{c}ek,
  ``Improved tensor multiplets,''
  Phys.\ Lett.\  B {\bf 109}, 439 (1982).

\bibitem{Siegel}
  W.~Siegel,
  ``Gauge spinor superfield as a scalar multiplet,''
  Phys.\ Lett.\  B {\bf 85}, 333 (1979).

\bi{rsg}
G.~Sierra and P.~K.~Townsend,
 ``An introduction to N=2 rigid supersymmetry,''
in {\it Supersymmetry and Supergravity 1983}, 
B. Milewski (Ed.), World Scientific, Singapore, 1983;
B.~de Wit, P.~G.~Lauwers, R.~Philippe, S.~Q.~Su and A.~Van Proeyen,
 ``Gauge and matter fields coupled to N = 2 supergravity,''
  Phys.\ Lett.\ B {\bf 134}, 37 (1984);
S.~J.~Gates, Jr., 
 ``Superspace formulation of new nonlinear sigma models,''
  Nucl.\ Phys.\ B {\bf 238}, 349 (1984).

\bibitem{DKT}
  N.~Dragon, S.~M.~Kuzenko and U.~Theis,
  ``The vector-tensor multiplet in harmonic superspace,''
  Eur.\ Phys.\ J.\  C {\bf 4}, 717 (1998)
  [arXiv:hep-th/9706169].

\bibitem{vev}
  I.~L.~Buchbinder, E.~I.~Buchbinder, E.~A.~Ivanov, 
S.~M.~Kuzenko and B.~A.~Ovrut,
 ``Effective action of the N = 2 Maxwell 
multiplet in harmonic superspace,''
Phys.\ Lett.\ B {\bf 412}, 309  (1997)
{[hep-th/9703147]};
I.~L.~Buchbinder and S.~M.~Kuzenko,
 ``The off-shell massive hypermultiplets revisited,''
 Class.\ Quant.\ Grav.\  {\bf 14}, L157  (1997)
{[hep-th/9704002]};
 N.~Dragon and S.~M.~Kuzenko,
 ``The Higgs Mechanism in N = 2 superspace,''
Nucl. Phys. B {\bf 508},  229  (1997)
{hep-th/9705027};
E.~A.~Ivanov, S.~V.~Ketov and B.~M.~Zupnik,
 ``Induced hypermultiplet self-interactions in N = 2 gauge theories,''
Nucl.\ Phys.\ B {\bf 509}, 53  (1998) 
{[hep-th/9706078]}.

\bibitem{projective3}
  F.~Gonzalez-Rey and R.~von Unge,
  ``Feynman rules in N = 2 projective superspace. II: Massive
  hypermultiplets,''
  Nucl.\ Phys.\ B {\bf 516}, 449 (1998) 
  {[hep-th/9711135]}.

\bibitem{K06}
  S.~M.~Kuzenko,
  ``On superpotentials for nonlinear sigma-models with eight supercharges,''
  Phys.\ Lett.\  B {\bf 638}, 288 (2006)
  [arXiv:hep-th/0602050].

\bibitem{BILS}
I.~A.~Bandos, E.~Ivanov, J.~Lukierski and D.~Sorokin,
``On the superconformal flatness of AdS superspaces,''
JHEP {\bf 0206}, 040 (2002) [hep-th/0205104].

\bibitem{HT}
P.~S.~Howe and R.~W.~Tucker,
``Scale invariance in superspace,''
Phys.\ Lett.\  B {\bf 80}, 138 (1978).

\bibitem{GSW}
 R.~Grimm, M.~Sohnius and J.~Wess,
  ``Extended supersymmetry and gauge theories,''
Nucl.\ Phys.\  B {\bf 133}, 275 (1978).

\bibitem{Ferber}
  A.~Ferber, ``Supertwistors and conformal supersymmetry,''
  Nucl.\ Phys.\ B {\bf 132}, 55 (1978).

\bibitem{Park}
  J.~H.~Park,
  ``Superconformal symmetry and correlation functions,''
  Nucl.\ Phys.\ B {\bf 559}, 455 (1999)   [hep-th/9903230].
 
 \bibitem{KT}
 S.~M.~Kuzenko and S.~Theisen,
  ``Correlation functions of conserved currents in N = 2 superconformal
  theory,''
  Class.\ Quant.\ Grav.\  {\bf 17}, 665 (2000)
  [hep-th/9907107]; 
   I.~L.~Buchbinder, S.~M.~Kuzenko and A.~A.~Tseytlin,
  ``On low-energy effective actions in N = 2,4 superconformal theories in  four
  dimensions,''  Phys.\ Rev.\  D {\bf 62}, 045001 (2000) [hep-th/9911221]. 

\bi{Sohnius} M.~F.~Sohnius,
 ``Supersymmetry and central charges,''
  Nucl.\ Phys.\ B {\bf 138}, 109 (1978).

\bibitem{Fayet}
P.~Fayet, ``Fermi-Bose hypersymmetry,''
Nucl.\ Phys.\ B {\bf 113}, 135 (1976).

\bibitem{West}
  P.~C.~West,
  {\it Introduction to supersymmetry and supergravity},
World Scientific, Singapore, 1990. 

\bibitem{KTM4}
  S.~M.~Kuzenko and G.~Tartaglino-Mazzucchelli,
  ``Conformally flat supergeometry in five dimensions,''
  arXiv:0804.1219 [hep-th].

\bibitem{BBKO}
  I.~L.~Buchbinder, E.~I.~Buchbinder, S.~M.~Kuzenko and B.~A.~Ovrut,
  ``The background field method for N = 2 super Yang-Mills theories in
  harmonic superspace,''
  Phys.\ Lett.\  B {\bf 417}, 61 (1998)
  [hep-th/9704214].

\bibitem{BKO}
  I.~L.~Buchbinder, S.~M.~Kuzenko and B.~A.~Ovrut,
  ``On the D = 4, N = 2 non-renormalization theorem,''
  Phys.\ Lett.\  B {\bf 433}, 335 (1998)
  [hep-th/9710142].

\bibitem{Grimm}
R.~Grimm,
``Solution of the Bianchi identities in SU(2) extended superspace with
constraints,''
in {\it Unification of the Fundamental Particle Interactions}, 
S. Ferrara, J. Ellis and P. van Nieuwenhuizen (Eds.),
Plenum Press, New York, 1980,
pp. 509-523.
      



\end{thebibliography}
\end{document}